\RequirePackage{fix-cm}
\documentclass[twocolumn]{svjour3}          
\smartqed  
\usepackage{graphicx}
\usepackage{amssymb}
\usepackage{amsmath}
\usepackage{epstopdf}
\usepackage[sort&compress,comma]{natbib}
\usepackage{booktabs}
\usepackage{siunitx}
\usepackage{soul}
\usepackage[mathscr]{euscript}
\usepackage{comment}
\usepackage{ulem}

\usepackage{pgfplots}
\pgfplotsset{compat=newest}

\usepackage{color}

\usepackage{xspace}
\newcommand{\ie}{i.\,e.\xspace}
\newcommand{\eg}{e.\,g.\xspace}

\newcommand{\iek}{i.\,e., }

\begin{document}

\title{
Investigating the spatial resolution of EMG and MMG based on a systemic multi-scale model 
}
\subtitle{}


\author{Thomas Klotz \and
        Leonardo Gizzi \and
        Oliver R\"ohrle
        }


\institute{T.~Klotz$^{*}$ \and L.~Gizzi \and O.~R\"ohrle \at
              Institute for Modelling and Simulation of Biomechanical Systems, Pfaffenwaldring 5a, 70569 Stuttgart, Germany \\
              Stuttgart Centre for Simulation Science (SimTech), Pfaffenwaldring 5a, 70569 Stuttgart, Germany\\
              Tel.: +49 711 685 66216 \\
        \and
           T.~Klotz$^{*}$ \at \email{thomas.klotz@imsb.uni-stuttgart.de}           
        \and
           L.~Gizzi \at \email{leonardo.gizzi@imsb.uni-stuttgart.de}   
        \and
           O.~R\"ohrle \at \email{roehrle@simtech.uni-stuttgart.de}      
}

\date{Received: date / Accepted: date}

\sloppy
\maketitle

\begin{abstract}
While electromyography (EMG) and magnetomyography (MMG) are both methods to measure the electrical activity of skeletal muscles, no systematic comparison between both signals exists.
Within this work, we propose a systemic \textit{in silico} model for EMG and MMG and test the hypothesis that MMG surpasses EMG in terms of spatial selectivity. 
The results show that MMG provides a slightly better spatial selectivity than EMG when recorded directly on the muscle surface.
However, there is a remarkable difference in spatial selectivity for non-invasive surface measurements.
The spatial selectivity of the MMG components aligned with the muscle fibres and normal to the body surface outperforms the spatial selectivity of surface EMG.
Particularly, for the MMG's normal-to-the-surface component the influence of subcutaneous fat is minimal.
Further, for the first time, we analyse the contribution of different structural components, \ie, muscle fibres from different motor units and the extracellular space, to the measurable biomagnetic field.
Notably, the simulations show that the normal-to-the-surface MMG component, the contribution from volume currents in the extracellular space and in surrounding inactive tissues is negligible. 
Further, our model predicts a surprisingly high contribution of the passive muscle fibres to the observable magnetic field.
\end{abstract}

\section{Introduction}\label{intro}
Movement relies on the complex interplay of the neural and musculoskeletal system. 
In short, the neuromuscular system comprises motor units, consisting of a motor neuron and all muscle fibres it innervates \citep{Heckman2012}.
Motor neurons integrate signals from the brain, sensory organs and recurrent pathways.
Once a motor neuron surpasses its depolarisation threshold, it triggers an action potential that propagates along the respective axon to the neuromuscular junctions.
The latter, opens ion channels in the sarcolemma, \ie, the muscle fibre membrane, yielding an action potential that travels along the muscle fibre triggering an intracellular signalling cascade that ultimately leads to force production, cf. \eg, \citet{MacIntosh2006,Roehrle2019}. 

From a physical point of view, an action potential represents a coordinated change of a membrane's polarity and thus causes both a time-dependent electric field, \ie, due to the distribution of charges, and a magnetic field, \ie, due to the flux of charges.
This can exploited for observing a skeletal muscle's activity via electromyography (EMG) or magnetomyography (MMG).    
Both signals contain information on the neural drive to the muscle and the state of the muscle, and, thus, can be both utilised to investigate various aspects of neuromuscular physiology. 
In the past, however, it was almost only EMG that has been used to study neuromuscular physiology (for a review see \cite{Merletti2016}). 
While EMG can be recorded either intramuscularly or from the body surface, from a practical point of view, non-invasive measurements are desirable. 
Signals obtained from surface EMG, however, exhibit limited spatial resolution, as the volume conductive properties of subcutaneous tissues act as low pass filter.
This means, a single surface EMG channel records from relatively large tissue volumes making it challenging to separate and accurately reconstruct the bioelectromagentic sources.
As the magnetic permeability of biological tissues is close to the magnetic permeability in free space \citep{Malmivuo1995,Oschman2002}, MMG has the potential to outperform the spatial resolution of surface EMG.
Further, in contrast to EMG, MMG recordings do not rely on sensor-tissue contacts and thus are particularly appealing for long term measurements; \eg, prosthesis control via implanted sensors \citep{Zuo2020}.   
Although MMG was already first described by \citeauthor{Cohen1972} in \citeyear{Cohen1972}, there still exist several challenges that limit its practical use.
Most importantly the amplitude of the magnetic field induced by skeletal muscles is very low, \ie, in the range of pico- to femto-Tesla and, thus, significantly lower than the earth's magnetic field.
This yields high technical demands for MMG recording systems \citep{Zuo2020}, for example, with respect to the sensitivity, the detection range, the sampling rate, the shielding from magnetic noise, the size and portability of the sensor device as well as the cost of such recordings.     
Nevertheless, a few proof-of-concept studies, \eg \citet{Reincke1993,Broser2018,Llinas2020,Broser2021}, illustrate its feasibility for biomedical applications.  

Despite originating from the same phenomenon, there hardly exist any studies that investigate the biophysical factors affecting MMG or compare MMG recordings with EMG.
Beside experimental studies, systemic \textit{in silico} models can be used to investigate the factors influencing bioelectromagentic signals and to test hypothesis derived from experimental observations.
Particularly, continuum field models have been successfully used for assisting the interpretation of EMG signals, \eg, \citet{Farina2002,Mesin2005,Dimitrova1999,Lowery2002,Farina2004b,Mesin2006a,Mordhorst2015,Mordhorst2017,Klotz2020}.
In contrast, models to simulate magnetic fields induced by skeletal muscles are still rare.
Common to all MMG models is that they first calculate the current field, which is then used to obtain the magnetic field. 
For example, \cite{Broser2021} used a finite wire model to infer from their measurements the underling physiology.
However, this approach could not explain some of their experimental observations. 
This is mainly due to the oversimplification of the muscle's anatomy as well as its physiology.     
\citet{Zuo2020,Zuo2021} followed a full-field approach, which was originally proposed by \cite{Woosley1985}, to simulate the magnetic field of an isolated axon. 
Thereby, the muscle fibres and the extracellular connective tissue are modelled as spatially separated regions, whereby the coupling conditions are determined from a pre-computed transmembrane potential.
While this approach allows to calculate both the electrical potential field and the magnetic field in a small tissue sample, the computational demands are substantial limiting its use for simulating larger tissue samples. 
Further, the decoupling of the transmembrane potential from the intracellular and extracellular potential fields is a simplification potentially limiting the credibility of the resulting modelling predictions.

To enable systematic in silico investigations for both EMG and MMG signals, we extend our homogenised multi-domain modelling framework \citep{Klotz2020} to predict both the skeletal-muscle-induced electric and magnetic field.
After establishing the model, we first investigate the hypothesis that for non-invasive recordings MMG provides a better spatial selectivity than EMG. 
Further, we use our model to quantify the contributions of different structural components to a muscle induced biomagentic field.

\section{Methods}
%
\begin{figure*}[ht!]
  \center
  \includegraphics[width=1.0\textwidth]{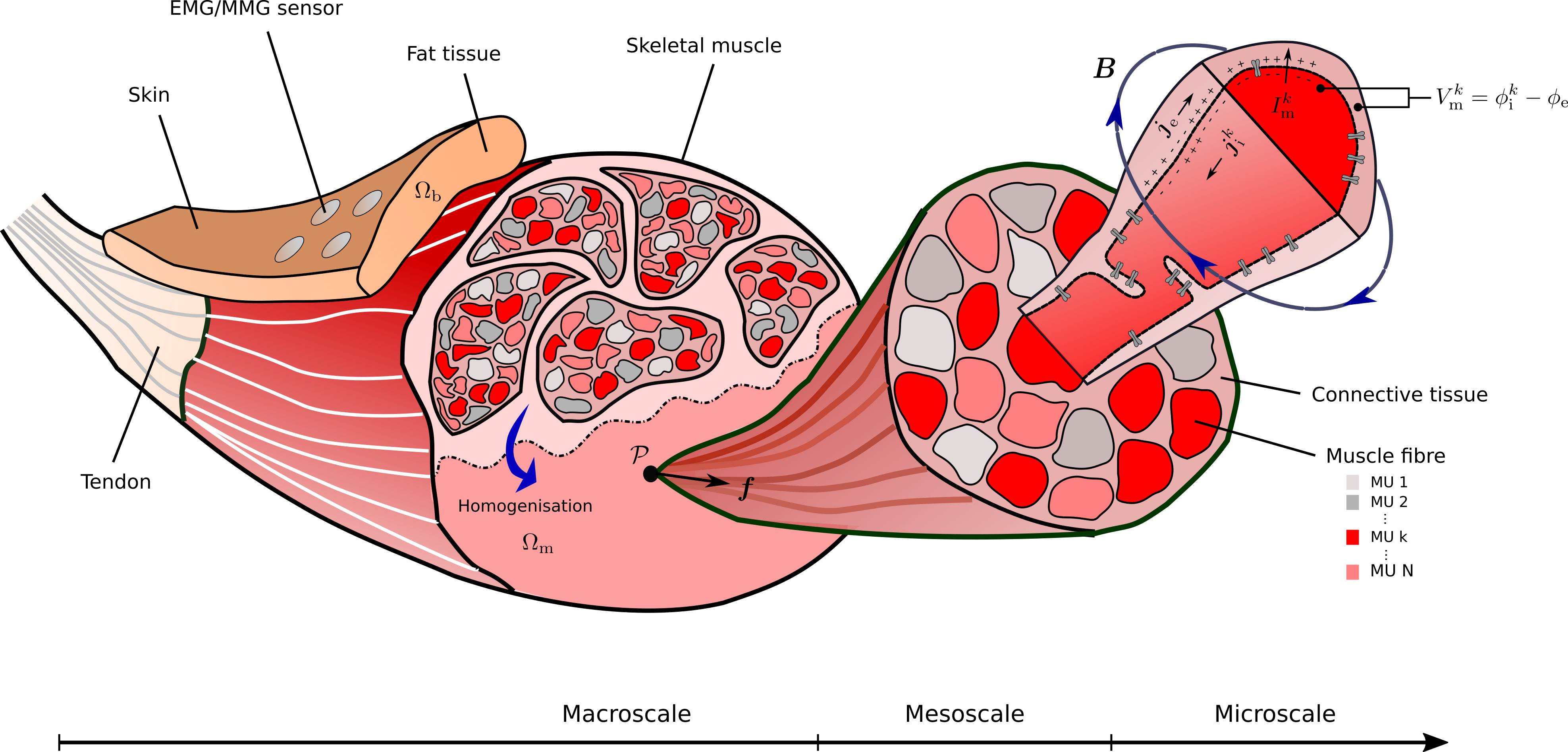}
  \caption[]{Schematic drawing illustrating the concept of the proposed multi-scale model. 
  On the macroscale, the heterogeneous muscle structure is smeared and represented by an idealised and continuous multi-domain material.  
  To couple the different domains, the multi-scale model still captures the most important features of the original structure.
  That is, the motor unit composition on the mesoscale and the interaction of one representative muscle fibre per motor unit and the extracellular space  through the muscle fibre membrane on the microscale.
  Note that for the muscle fibres each colour represents a different motor unit.
  Based on those key properties, the continuous field approach predicts experimental measurable fields such as the transmembrane potentials $V_\mathrm{m}^k$, the extracellular potential $\phi_\mathrm{e}$ or the magnetic $\boldsymbol{B}$ field.
  Particularly non-invasive surface recordings, which are schematically illustrated by the sensors on the surface, of the electrical potential field, \ie, via electromyography (EMG), or the magnetic field, \ie, via magnetomyography (MMG), are preferable as they yield minimal discomfort for a subject. 
  }\label{fig:graphical_abstract}
\end{figure*}
\subsection{Modelling framework}\label{sec:modelling_framework}
This section presents the modelling framework for investigating relations between the biophysical state of the neuromuscular system and muscle induced bioelectromagnetic fields.
The underlying governing equations are the quasi-static Maxwell's equations as presented in Sect.~\ref{sec:maxwells_eqs}.
As the quasi-static approximation of Maxwell's equations allows us to decouple the electric field from the magnetic field, we first derive a systemic multi-scale model to simulate the electro-physiological behaviour of skeletal muscles \citep[][cf. Sect.~\ref{sec:multi_domain_muscle}]{Klotz2020} as well as electrically inactive tissue that surrounds the respective muscle tissue (cf. Sect.~\ref{sec:electrically_inactive_tissues}).
Based on the electric potential field in the region of interest, the corresponding current densities, and, hence, the prediction of the magnetic field, can be calculated (cf. Sect.~\ref{sec:modelling_magnetic_field}). 
Sect.~\ref{sec:boundary_conditions} provides appropriate boundary conditions to guarantee existence and uniqueness for the solution of the derived system of partial differential equations.  
\subsubsection{Governing equations}\label{sec:maxwells_eqs}
In classical physics, the evolution of the electric and magnetic field is described by Maxwell's equations. 
Since changes to the muscle induced electric and magnetic field are relatively slow, \ie, the characteristic time scale is in the range of milliseconds, the electrostatic and the magnetostatic approximation holds for modelling the EMG and MMG signal.
The differential form of the quasi-static Maxwell's equations is given by, \eg, \citet{Griffiths2013},
\begin{subequations}
\begin{align}
  \operatorname{div} \boldsymbol{E} \ &= \ \frac{v}{\varepsilon_0}, & (\text{Gauss's law})  \label{eq:static_gauss_law} \\ 
  \operatorname{div} \boldsymbol{B} \ &= \ 0, & (\text{Gauss's law for magnetism}) \label{eq:static_gauss_law_magnetism} \\ 
  \quad {\operatorname{curl}}\ \boldsymbol{E} \ &= \ \boldsymbol{0}  , & (\text{Faraday's law}) \label{eq:static_maxwell_faraday}\\ 
  \quad {\operatorname{curl}}\ \boldsymbol{B} \ &= \ \mu_0 \, \boldsymbol{j}.  & (\text{Amp\`{e}re's law}) \label{eq:static_amperes_law}
\end{align}
\end{subequations}
Therein, $\operatorname{div}(\cdot)$ denotes the divergence operator, ${\operatorname{curl}}(\cdot)$ denotes the curl operator, $\boldsymbol{E}$ is the electrical field intensity, $v$ is the electrical charge density, $\varepsilon_0$ is the vacuum permittivity, $\boldsymbol{B}$ is the magnetic field (sometimes also referred to as magnetic flux density), $\mu_0$ is the vacuum permeability and $\boldsymbol{j}$ is the total electrical current density.
Further, applying the div-curl identity to Amp\`{e}re's law (Eqn.~\eqref{eq:static_amperes_law}) yields the conservation of charges, \ie,
\begin{equation}\label{eq:static_charge_balance}
  \operatorname{div} \boldsymbol{j} \ = \ 0 \ . 
\end{equation}
Exploiting the fact that the electrical field intensity, $\boldsymbol{E}$, is a conservative vector field and, thus, can be derived from a scalar potential, \ie $\boldsymbol{E} = - \operatorname{grad} \phi$ with $\operatorname{grad}(\cdot)$ being the gradient operator, reduces the number of state variables.
Further, introducing the magnetic vector potential $\boldsymbol{A}$ such that
\begin{equation}\label{eq:mag_vec_pot_def}
  \boldsymbol{B} = {\operatorname{rot}}\ \boldsymbol{A}
\end{equation}   
and calibrating it by Coulomb gauge, \ie, $\operatorname{div} \ \boldsymbol{A} = 0$, we obtain the quasi-static Maxwell's equations in potential form:  
\begin{subequations}
\begin{align}
  \operatorname{div}(\operatorname{grad} \phi) \ &= \ - \frac{v}{\varepsilon_0} \ , \label{eq:static_maxwell_faraday_pot} \\
  \operatorname{div}(\operatorname{grad} \boldsymbol{A}) \ &= \ - \mu_0 \, \boldsymbol{j} \ . \label{eq:static_amperes_law_pot}
\end{align}
\end{subequations} 
Next, we will introduce suitable modelling assumptions reflecting the electro-physiological properties of skeletal muscle tissue.
Thereby note that for skeletal muscles, bound currents are assumed to be negligible and thus the total current density $\boldsymbol{j}$ is equal to the ''free'' current density (which is also sometimes called conductive current density).  

\subsubsection{Modelling the electrical behaviour of skeletal muscles}\label{sec:multi_domain_muscle}
The electrical behaviour of skeletal muscles is simulated based on the multi-domain model presented in \cite{Klotz2020} and is briefly summarised here.
Skeletal muscle tissue consists of muscle fibres associated with different motor units and extracellular connective tissue (cf. Fig.~\ref{fig:graphical_abstract}).
The multi-domain model resolves this tissue heterogeneity by assuming that there coexist at each skeletal muscle material point $\mathcal{P}\in \mathrm{\Omega_m}$ an extracellular space and $N$ intracellular spaces, with $N$ denoting the number of motor units.
Given this homogenized tissue representation, an electrical potential is introduced for each domain, \iek $\phi_\mathrm{e}$ and $\phi_\mathrm{i}^k$, $\forall \ k \in \mathscr{M}_\mathrm{MU}:=\{1,2,...,N \}$, where the subscripts $(\cdot)_\mathrm{e}$ and  $(\cdot)_\mathrm{i}$ denote extracellular and intracellular quantities, respectively. 
Further, a transmembrane potential $V_\mathrm{m}^k$ is introduced for each motor unit, \ie, 
\begin{equation}\label{eq:def_vm}
	V_\mathrm{m}^k=\phi_\mathrm{i}^k-\phi_\mathrm{e} \ , \ \forall \ k \in \mathscr{M}_\mathrm{MU} \ .
\end{equation}

The domains are electrically coupled, which is modelled by taking into account the most important features of skeletal muscles mesostructure and microstructure as well as the dynamics of the muscle fibre membranes.  
Thus, the multi-domain model can be classified as a multi-scale model.

The conservation of charges, \ie, Eqn.\eqref{eq:static_charge_balance}, requires that all outward volume fluxes of the current densities from all domains are balanced at each skeletal muscle material point.
For skeletal muscles it can be assumed that ions can only be exchanged between an intracellular domain and the extracellular space.
There exist no current fluxes between the different intracellular domains.
As the muscle fibres of the same motor unit are assumed to show similar biophysical properties, the coupling between an intracellular space $k$ and the extracellular space is modelled by considering the interaction of one representative muscle fibre per motor unit with the extracellular space.   
Therefore, the current density outward volume flux of an intracellular domain is
\begin{equation}\label{eq:balance_int}
  \operatorname{div} \boldsymbol{j}_\mathrm{i}^k \ = \ A_\mathrm{m}^k I_\mathrm{m}^k  \ , \ k \in \mathscr{M}_\mathrm{MU} \ , \quad  \text{in} \ \mathrm{\Omega_m} \, ,
\end{equation}
where $\boldsymbol{j}_\mathrm{i}^k$ is the current density of motor unit $k$ in a representative fibre-matrix cylinder. 
Further, $A_\mathrm{m}^k$ is the surface-to-volume ratio of a muscle fibre belonging to motor unit $k$, \ie, representing the geometry of the muscle fibres on the microscale, and $I_\mathrm{m}^k$ is the transmembrane current density, \ie, resolving the (microscale) behaviour of the muscle fibre membranes. 
The conservation of charges holds for each skeletal muscle material point if the current density outward volume flux from the extracellular domain is equal to the weighted sum of the transmembrane current densities, \ie,
\begin{equation} \label{eq:balance_ext} 
  \operatorname{div} \boldsymbol{j}_\mathrm{e} \ = \ - \sum_{k=1}^N \, f_\mathrm{r}^k A_\mathrm{m}^k I_\mathrm{m}^k \ , \quad \text{in} \ \mathrm{\Omega_m} \, ,
\end{equation}
where $\boldsymbol{j}_\mathrm{e}$ is the extracellular current density.
Further, $f_\mathrm{r}^k$ is a (mesoscale) parameter, reflecting the motor unit composition at each skeletal muscle material point, \ie the volume fraction of all muscle fibres belonging to motor unit $k$ ($\forall k \ \in \mathscr{M}_\mathrm{MU}$) divided by the volume fraction of all muscle fibres.

The (conductive) current densities are related to the electrical potential fields via Ohm's law, \ie,  
\begin{equation}\label{eq:Ohms_law}
\begin{aligned}
  \boldsymbol{j}_\mathrm{e} \ &= \ -\boldsymbol{\sigma}_\mathrm{e} \operatorname{grad} \phi_\mathrm{e} \ , \\
  \boldsymbol{j}_\mathrm{i}^k \ &= \ -\boldsymbol{\sigma}_\mathrm{i}^k \operatorname{grad} \phi_\mathrm{i}^k \ , \ \forall \ k \in \mathscr{M}_\mathrm{MU} \ ,
\end{aligned}
\end{equation}
where $\boldsymbol{\sigma}_\mathrm{e}$ and $\boldsymbol{\sigma}_\mathrm{i}^k$ denote the extracellular conductivity tensor and the intracellular conductivity tensors, respectively.

Finally, the transmembrane current densities, $I_\mathrm{m}^k$ ($\forall k \ \in \mathscr{M}_\mathrm{MU}$), are calculated from an electrical circuit model \citep{Hodgkin1952,Keener2009b} of the muscle fibre membranes via Kirhhoff's current law, \ie, 
\begin{equation}\label{Eq:Hodgkin_huxley_membrane}
\begin{aligned}
    I_\mathrm{m}^k  \ &= \ C_\mathrm{m}^k \dot{V}_\mathrm{m}^k \, + \, I_\mathrm{ion}^k(\boldsymbol{y}^k,V_\mathrm{m}^k, I_{\mathrm{stim}}^k)  \ , \\
   \dot{\boldsymbol{y}}^k\ &= \ \boldsymbol{g}^k(\boldsymbol{y}^k,V_\mathrm{m}^k) \ , \\
   \boldsymbol{y}_0^k \ &= \ \boldsymbol{y}^k(t=0) \ , \\ 
	V_\mathrm{m,0}^k \ &= \ V_\mathrm{m}^k(t=0) \ . \\
\end{aligned}  
\end{equation}
Therein, $C_\mathrm{m}^k$ is the membrane capacitance per unit area of a muscle fibre belonging to motor unit $k$, $I_\mathrm{ion}^k(\boldsymbol{y}^k(t),V_\mathrm{m}^k, I_{\mathrm{stim}}^k)$ is the total ohmic current density through a membrane patch associated with MU $k$ and $I_{\mathrm{stim}}^k$ is an external stimulus that is used to describe the motor nerve stimuli of motor unit $k$ at the neuromuscular junctions.
Further, $\boldsymbol{y}^k$ is a vector of additional state variables, \eg, describing the probability of ion channels to be open or closed and $\boldsymbol{g}^k(\boldsymbol{y}^k,V_\mathrm{m}^k)$ is a vector-valued function representing the evolution equation for the membrane state vector $\boldsymbol{y}^k$.    

Combing Eqns.~\eqref{eq:def_vm}-%
\eqref{Eq:Hodgkin_huxley_membrane} yields for each 
$\mathcal{P}\in \mathrm{\Omega_m}$ the following system of coupled differential equations:
\begin{subequations} 
\begin{align}
  0 \ &= \ \operatorname{div} \left[\boldsymbol{\sigma}_\mathrm{e} \operatorname{grad} \phi_\mathrm{e} \right] \, \notag \\ 
  & \quad \ \ + \, \sum_{k=1}^N  \, f_\mathrm{r}^k \operatorname{div}
  \left[\boldsymbol{\sigma}_\mathrm{i}^k \operatorname{grad} \left(V_\mathrm{m}^k + \phi_\mathrm{e} \right) \right] \ , \label{extracellular_md_eqs}  \\
  \displaystyle\frac{\partial V_\mathrm{m}^k}{\partial t} \ &= \
  \frac{1}{C_\mathrm{m}^k A_\mathrm{m}^k} \Big( \operatorname{div} \, \left[\boldsymbol{\sigma}_\mathrm{i}^k \operatorname{grad} \left(V_\mathrm{m}^k  \phi_\mathrm{e} \right) \right]  & \notag \\ 
   & \quad \ \ - \, A_\mathrm{m}^k I_\mathrm{ion}^k(\boldsymbol{y}^k,V_\mathrm{m}^k) \Big) \ , \ \forall \ k \in \mathscr{M}_\mathrm{MU} \ , \label{intracellular_md_eqs} \\
   \dot{\boldsymbol{y}}^k \ &= \ \boldsymbol{g}^k(\boldsymbol{y}^k,V_\mathrm{m}^k) \ , \ \qquad \qquad \forall \ k \in \mathscr{M}_\mathrm{MU} \ . \label{intracellular_md_eqs_odes}
\end{align}
\end{subequations}
Further details can be found in \citet{Klotz2020}.

\subsubsection{Modelling the electrical behaviour of inactive tissues}\label{sec:electrically_inactive_tissues}
Skeletal muscles are surrounded by electrically inactive tissues, \eg, connective tissues, fat or skin.
Electrically inactive tissues have a strong influence on the electrical potential on the body surface.
From a modelling point of view, inactive tissue is a volume conductor free of current sources, cf., \eg, \citet{Pullan2005,Mesin2013} or \citet{Klotz2020}, yielding a generalised Laplace equation for each material point within the body region $\mathcal{P}\in \mathrm{\Omega_b}$, \ie,
\begin{equation}\label{eq:body_pot}
  \operatorname{div} \left[ \boldsymbol{\sigma}_\mathrm{b} \operatorname{grad} \phi_\mathrm{b} \right] \ = \ 0 \ , \quad \text{in} \ \mathrm{\Omega_b} \ ,
\end{equation}
where $\phi_\mathrm{b}$ and $\boldsymbol{\sigma}_\mathrm{b}$ are the  body region's electrical potential and conductivity tensor, respectively.

\subsubsection{Modelling magnetic fields induced by skeletal muscle's electrical activity}\label{sec:modelling_magnetic_field}
Starting point for predicting the magnetic field is Amp\`{e}re's law, \ie, Eqn.~\eqref{eq:static_amperes_law} or Eqn.~\eqref{eq:static_amperes_law_pot}, which relates the magnetic field to the total current density.
Exploiting that the radius of a muscle fibre is small compared to the characteristic length scale of the macroscopic continuum model and that the muscle fibres are (approximately) of cylindrical shape leads to the assumption that the contributions of the transmembrane currents to the macroscopic magnetic field cancel each other out. 
Further, assuming that for skeletal muscle tissue the magnetic susceptibility is approximately zero and that, within the limits of the qausi-static approximation, polarisation currents are negligible (cf. e.g., \citet{Malmivuo1995}), then the overall current density is fully determined by the conductive current densities.
The latter is related to the electrical potential field via Ohm's law, and thus, for the homogenised multi-domain model the current density can be calculated for each domain independently (cf. Eqn.~\eqref{eq:Ohms_law}).
To derive the right-hand side of Amp\`{e}re's law from domain-specific current densities, we consider its integral form, \ie,
\begin{equation}\label{eq:int_amperes_law}
  \oint_C \boldsymbol{B} \, \mathrm{d}\boldsymbol{l} \ = \ \mu_0 I_\mathrm{enc} \ .
\end{equation}
Therein $C$ is an arbitrary closed curve, $I_\mathrm{enc}$ is the total current crossing $C$ and $\mathrm{d}\boldsymbol{l}$ is an infinitesimal line element. 
Eqn.~\ref{eq:int_amperes_law} shows that the currents of the individual domains simply add up linearly.
Since the current densities are given with respect to a representative fibre-matrix cylinder, the contributions of the intracellular current densities $\boldsymbol{j}_\mathrm{i}^k$ ($\forall k \in \mathscr{M}_\mathrm{MU}$) need to be weighted by the (mesoscale) motor unit density factor $f_\mathrm{r}^k$. 
Accordingly, the magnetic vector potential for every material point within the muscle region $\mathcal{P} \in \mathrm{\Omega_m}$ is
\begin{equation}\label{eq:mag_pot_muscle}
\begin{aligned}
\operatorname{div}(\operatorname{grad}  \boldsymbol{A}_\mathrm{m}) \ &=\ \mu_0  \,
  \Big(
    \boldsymbol{j}_\mathrm{e}  \, + \, \sum_{k=1}^N  f_\mathrm{r}^k \boldsymbol{j}_\mathrm{i}^k 
  \Big)  \ , \\
  \Leftrightarrow
  \operatorname{div}(\operatorname{grad}  \boldsymbol{A}_\mathrm{m}) \ &=\ - \mu_0  \,
  \Big(
    \boldsymbol{\sigma}_\mathrm{e} \operatorname{grad} \phi_\mathrm{e} \\
    &+ \, \sum_{k=1}^N  f_\mathrm{r}^k \boldsymbol{\sigma}_\mathrm{i}^k \operatorname{grad} (V_\mathrm{m}^k + \phi_\mathrm{e})
  \Big)  \ .
\end{aligned}
\end{equation}
Note, the potential formulation is chosen as this yields a Poisson-type equation for which various well-established numerical solution methods exist.
Further, note that the linearity of the magnetostatic equations can be exploited to predict the contribution of each domain to the experimentally observable magnetic field.
The body's magnetic vector potential, $\boldsymbol{A}_\mathrm{b}$,  is calculated similarly:
\begin{equation}\label{eq:mag_pot_body}
\begin{aligned}
  \operatorname{div}(\operatorname{grad}  \boldsymbol{A}_\mathrm{b}) \ &= \ \mu_0 \, \boldsymbol{j}_\mathrm{b} \ , \\
  \Leftrightarrow \operatorname{div}(\operatorname{grad}  \boldsymbol{A}_\mathrm{b}) \ &= \ - \mu_0  \, \left[ \boldsymbol{\sigma}_\mathrm{b} \operatorname{grad} \phi_\mathrm{b} \right] \ , 
\end{aligned}
\end{equation}
where $\boldsymbol{j}_\mathrm{b}$ is the current density in the body region.
In contrast to the electrical field equations, the magnetic field equations also need to consider the air surrounding the body. 
Since air can be  assumed to be free of electrical currents, it is modelled by 
\begin{equation}\label{eq:mag_pot_free}
  \operatorname{div}(\operatorname{grad}  \boldsymbol{A}_\mathrm{f}) \ = \ \boldsymbol{0} \ , \ \text{in} \ \mathrm{\Omega_f} \ ,
\end{equation}
where $\boldsymbol{A}_\mathrm{f}$ is the magnetic vector potential within the surrounding space $\mathrm{\Omega_f}$. 

Finally, the experimentally measurable magnetic field $\boldsymbol{B}$ can be calculated straight forwardly from Eqn.\eqref{eq:mag_vec_pot_def}.

\subsubsection{Boundary conditions}\label{sec:boundary_conditions}
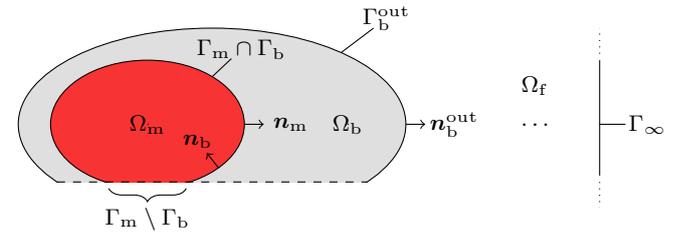
\begin{figure}[h]
	\center
  \begin{minipage}{0.8\textwidth}
    \begin{tikzpicture}[scale=0.85]
\definecolor{cf6d1d1}{RGB}{246,0,0}
\definecolor{cc58888}{RGB}{246,0,0}

\begin{scope}
\clip(-1.1cm,-0.9cm)rectangle(10cm,2cm);

\draw [black,fill=gray!25] (2cm,0cm) ellipse (3cm and 1.5cm);
\draw [black,fill=cc58888!80] (1cm,0cm) ellipse (1.5cm and 1.0cm);
\draw[opacity=1] (1cm,0cm) -- (1cm,0cm) node[pos=0]{$\mathrm{\Omega_m}$}; 
\draw[opacity=1] (4.1cm,0cm) -- (4.1cm,0cm) node[pos=0]{$\mathrm{\Omega_b}$};
\draw[opacity=1] (4.5cm,1.5cm) -- (4.0cm,1.12cm) node[pos=-0.4]{$\mathrm{\Gamma_b^{out}}$};
\draw[opacity=1] (2.3cm,1.025cm) -- (2.0cm,0.74cm) node[pos=-0.5]{$\mathrm{\Gamma_m} \cap \mathrm{\Gamma_b}$};
\draw[opacity=1,->] (2.5cm,0cm) -- (2.8cm,0cm) node[pos=2.4]{$\boldsymbol{n}_\mathrm{m}$};
\draw[opacity=1,->] (2.1cm,-0.68cm) -- (1.92cm,-0.45cm) node[pos=1.8]{$\boldsymbol{n}_\mathrm{b}$};
\draw[opacity=1,->] (5cm,0cm) -- (5.3cm,0cm) node[pos=2.5]{$\boldsymbol{n}_\mathrm{b}^\mathrm{out}$};

\draw[opacity=1] (7cm,0.6cm) -- (7cm,0.6cm) node[pos=0]{$\mathrm{\Omega_f}$};
\draw[opacity=1] (7cm,0cm) -- (7cm,0cm) node[pos=0]{$\hdots$};
\end{scope}

\draw[opacity=1] (8cm,-0.8cm) -- (8cm,1.0cm); 
\draw[opacity=1, dotted] (8cm,-0.9cm) -- (8cm,-1.3cm); 
\draw[opacity=1, dotted] (8cm,1.1cm) -- (8cm,1.5cm); 
\draw[opacity=1] (8cm,0cm) -- (8.4cm,0cm) node[pos=1.85]{$\mathrm{\Gamma}_\infty$}; 

\draw[opacity=1, dashed] (-0.4cm, -0.9cm) -- (4.4cm,-0.9cm);
\draw [decorate,decoration={brace,mirror,amplitude=5pt}] (0.4cm,-1cm) -- (1.6cm,-1cm) node [midway,yshift=-0.4cm] {$\mathrm{\Gamma_m} \setminus \mathrm{\Gamma_b}$};

\end{tikzpicture}
  \end{minipage}
	\caption[]{Schematic illustration of an arbitrary geometrical representation of muscle tissue $\mathrm{\Omega_m}$, the body region $\mathrm{\Omega_b}$, the surrounding space $\mathrm{\Omega_f}$, and its respective interfaces. Thereby, $\mathrm{\Gamma_m}$ denotes the muscle boundary with unit outward normal vector $\boldsymbol{n}_\mathrm{m}$, $\mathrm{\Gamma_b^{out}}$ is the body surface with unit outward normal vector $\boldsymbol{n}_\mathrm{b}^\mathrm{out}$ , $\mathrm{\Gamma_b}$ is an inner boundary of the body region with unit outward normal vector $\boldsymbol{n}_\mathrm{b}$ and $\mathrm{\Gamma}_\infty$ refers to the set of infinitely distant points.}\label{fig:bcs}
\end{figure}

Suitable boundary conditions are required to solve the partial differential equations presented in the previous sections.
Recalling that muscle fibres are electrically insulated by their membranes, it is assumed that no charges can leave the intracellular domains at their boundary. 
This is modelled by applying zero Neumann boundary conditions to the intracellular potential, \ie,
\begin{equation} \label{eq:bc_Vm}
\begin{aligned}
  \left[ \boldsymbol{\sigma}_\mathrm{i}^k \operatorname{grad} \phi_\mathrm{i}^k \right] \cdot \boldsymbol{n}_\mathrm{m} \ &= \ 0 \ ,  \quad \ \ \ \, \text{on} \ \mathrm{\Gamma_m} \ , \\
  \Leftrightarrow \left[ \boldsymbol{\sigma}_\mathrm{i}^k \operatorname{grad} V_\mathrm{m}^k \right] \cdot \boldsymbol{n}_\mathrm{m} \ &  \\  
  = \ - \ \big[ \boldsymbol{\sigma}_\mathrm{i}^k \operatorname{grad} & \, \phi_\mathrm{e} \big] \cdot \boldsymbol{n}_\mathrm{m} , \quad  \text{on} \ \mathrm{\Gamma_m} \ ,
\end{aligned}  
\end{equation}
where ''$\cdot$'' denotes the scalar product and $\boldsymbol{n}_\mathrm{m}$ is a unit outward normal vector at the muscle surface $\mathrm{\Gamma_m}$ (cf. Fig.~\ref{fig:bcs}).

Further, it is assumed that no charges can leave the body, yielding zero Neumann boundary conditions for the electrical potential in the body region, \ie,
\begin{equation}\label{eq:bc_phi_b}
  \left[ \boldsymbol{\sigma}_\mathrm{b} \operatorname{grad} \phi_\mathrm{b} \right] \cdot \boldsymbol{n}_\mathrm{b}^\mathrm{out} \ = \ 0 \ , \quad \text{on} \ \mathrm{\Gamma_b^{out}} \ .
\end{equation}
Therein, $\boldsymbol{n}_\mathrm{b}^\mathrm{out}$ denotes a unit outward normal vector of the body surface $\mathrm{\Gamma_b^{out}}$ (cf. Fig.~\ref{fig:bcs}).
In case that the outer surface of the simulated region is the skeletal muscle tissue's boundary (or part thereof), the same assumption holds -- however with  zero Neumann boundary conditions for the extracellular potential, \ie,
\begin{equation}\label{eq:bc_phi_e_out}
  \left[ \boldsymbol{\sigma}_\mathrm{e} \operatorname{grad} \phi_\mathrm{e} \right] \cdot \boldsymbol{n}_\mathrm{m} \ = \ 0 \ , \quad \text{on} \ \mathrm{\Gamma_m} \setminus \mathrm{\Gamma_b} \ .
\end{equation} 
While these are idealised cases typically not reflecting exact \textit{in vivo} conditions, it should be noted that this boundary condition is still useful as most in silico experiments are restricted to a particular region of interest. 

Finally, it is assumed that at the muscle-body interface, the extracellular potential $\phi_\mathrm{e}$, and the electrical potential of the body region $\phi_\mathrm{b}$ are continuous, \ie,
\begin{equation}\label{eq:bc_phi_continuous} 
  \phi_\mathrm{e} \ = \ \phi_\mathrm{b} \ ,  \quad \, \ \text{on}  \ \mathrm{\Gamma_m} \cap \mathrm{\Gamma_b}  \ . 
\end{equation} 
Further, the current flux between the extracellular space and the body region is balanced, yielding
\begin{equation}\label{eq:bc_phi_flux}  
  \big[ \boldsymbol{\sigma}_\mathrm{e} \operatorname{grad} \phi_\mathrm{e} \, - \, \boldsymbol{\sigma}_\mathrm{b} \operatorname{grad} \phi_\mathrm{b} \big] \cdot \boldsymbol{n}_\mathrm{m} \ = \ 0 \ , \ \text{on} \ \mathrm{\Gamma_m} \cap \mathrm{\Gamma_b}  \ .
\end{equation}
Note electrical potential fields are not unique, \ie, they can be shifted by an arbitrary scalar value.
To make the solution unique, one can mimic/simulate a grounding electrode at a boundary location.

For the magnetic vector potential it can be assumed that far away from the muscle, \ie, the bioelectromagnetic sources, the magnetic field vanishes.
Thus zero Dirichlet boundary conditions are applied to all infinitely distant points $\mathrm{\Gamma}_\infty$, \ie,
\begin{equation}\label{eq:zero_dirichlet_mag_pot}
  \boldsymbol{A} \ = \ \boldsymbol{0} \ , \ \text{on} \  \mathrm{\Gamma}_\infty \ . 
\end{equation} 
It can be shown that the magnetic vector potential is continuous at the interface between two media (cf. \eg, \cite{Griffiths2013}).
This is modelled by 
\begin{subequations}\label{eq:interface_mag_pot}
\begin{align} 
  \boldsymbol{A}_\mathrm{m} \ &= \  \boldsymbol{A}_\mathrm{b} \ , \ \text{on} \  \mathrm{\Gamma_m} \cap \mathrm{\Gamma_b} \ , \\
  \boldsymbol{A}_\mathrm{m} \ &= \  \boldsymbol{A}_\mathrm{f} \ , \ \text{on} \  \mathrm{\Gamma_m} \setminus \mathrm{\Gamma_b} \ , \\
  \boldsymbol{A}_\mathrm{b} \ &= \  \boldsymbol{A}_\mathrm{f} \ , \ \text{on} \  \mathrm{\Gamma_b^{out}} \ .
\end{align} 
\end{subequations}
Further, for biological tissues, surface currents are assumed to be negligible (\ie, they only exhibit volume conduction).
Accordingly, the fluxes of the magnetic vector potential across any boundary are balanced \citep{Griffiths2013}, \ie,
\begin{subequations}\label{eq:interface_flux_mag_pot}
\begin{align}
   \big[ \operatorname{grad} \boldsymbol{A}_\mathrm{b} \, - \, \operatorname{grad}\boldsymbol{A}_\mathrm{m} \big]& \cdot \boldsymbol{n}_\mathrm{m}  \ = \ \boldsymbol{0} \ , \ \text{on} \ \mathrm{\Gamma_m} \cap \mathrm{\Gamma_b}  \ ,  \\
  \big[ \operatorname{grad} \boldsymbol{A}_\mathrm{f} \, - \, \operatorname{grad}\boldsymbol{A}_\mathrm{m} \big]& \cdot \boldsymbol{n}_\mathrm{m}  \ = \ \boldsymbol{0} \ , \ \text{on} \ \mathrm{\Gamma_m} \setminus \mathrm{\Gamma_b}  \ , \\
   \big[ \operatorname{grad} \boldsymbol{A}_\mathrm{f} \, - \, \operatorname{grad}\boldsymbol{A}_\mathrm{b} \big]& \cdot \boldsymbol{n}_\mathrm{b}^\mathrm{out}  \ = \ \boldsymbol{0} \ , \ \text{on} \ \mathrm{\Gamma_b^{out}}   \ .
\end{align} 
\end{subequations}

\subsection{In silico experiments}\label{sec:in_silico_exp}
\begin{table*}[t!]
\begin{tabular}{ l l l l} 
  Parameter & Symbol & Value (slow to fast) & Reference \\
	\toprule
  Longitudinal intracellular conductivity & $\sigma_\mathrm{i}^\mathrm{l}$ & \SI{8.93}{\milli\siemens\per\centi\meter} & \cite{Bryant1969}\\  
  Transversal intracellular conductivity & $\sigma_\mathrm{i}^\mathrm{t}$ & \SI{0.0}{\milli\siemens\per\centi\meter} & cf. \cite{Klotz2020} \\  
  Longitudinal extracellular conductivity & $\sigma_\mathrm{e}^\mathrm{l}$ & \SI{6.7}{\milli\siemens\per\centi\meter} & \cite{Rush1963} \\  
  Transversal extracellular conductivity & $\sigma_\mathrm{e}^\mathrm{t}$ & \SI{3.35}{\milli\siemens\per\centi\meter} & cf. \cite{Klotz2020} \\  
  Fat conductivity & $\sigma_\mathrm{b}$ & \SI{0.4}{\milli\siemens\per\centi\meter} & \cite{Rush1963} \\  \midrule
	Membrane capacitance & $C_\mathrm{m}^k$ & \SI{1}{\micro\farad\per\square\centi\meter} & \cite{Hodgkin1952} \\
	Surface-to-volume ratio & $A_\mathrm{m}^k$ & \SI{500}{\per\centi\meter} & cf. \cite{Klotz2020} \\ 
	Motor unit density & $f_\mathrm{r}^k$ & Variable \\\midrule
  Magnetic permeability & $\mu_0$ & 
\end{tabular} 
\caption[]{Summary of model parameters.}\label{tab:parameters} 
\end{table*}

\begin{figure}[h]
	\center
  \includegraphics[width=0.45\textwidth]{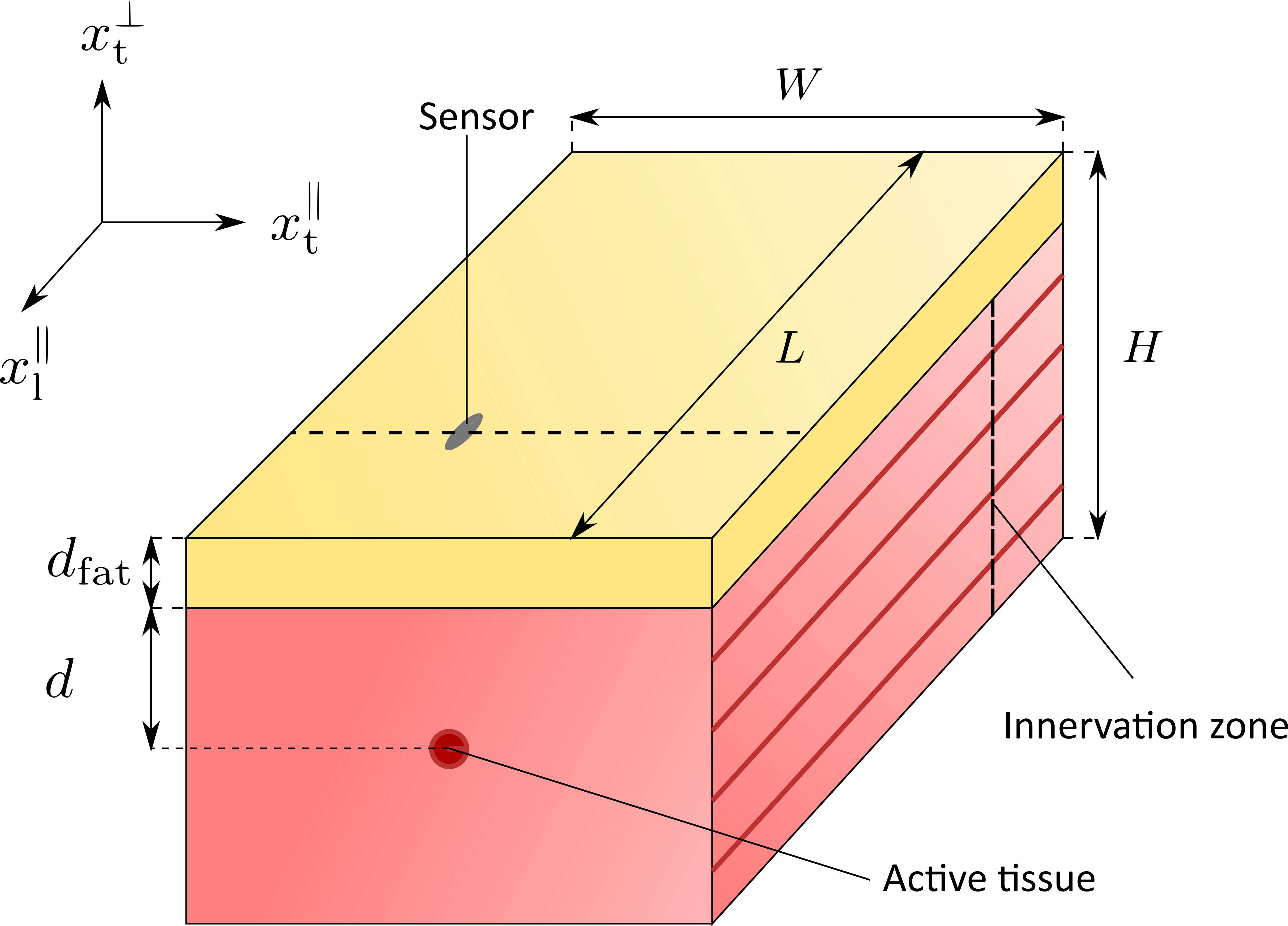}
	\caption[]{Schematic drawing illustrating the simulated tissue geometry, whereby muscle and fat tissue are coloured in red and yellow, respectively. The muscle fibres are aligned with the $x_\mathrm{l}^{\parallel}$-direction.}\label{fig:geo}
\end{figure}

The main aim of this work is to employ the previously described modelling framework to investigate the spatial resolution of non-invasive EMG and MMG.
This is achieved by simulating a muscle with a layer of subcutaneous fat on top and which is variable in thickness. 
We exclude the influence of the geometry by focusing on a cube-shaped (half) muscle sample with edge lengths $L=\SI{4.0}{\centi\meter}$, $W=\SI{1.5}{\centi\meter}$ and $H=\SI{2.0}{\centi\meter}$ (cf. Fig.~\ref{fig:geo}).  
The muscle fibres are aligned with the longest edge, \ie, denoted as the $x_\mathrm{l}^{\parallel}$-direction.
The spatial selectivity is addressed by a set of \textit{in silico} experiments, whereby the muscle fibres, \ie, the intracellular domains, are selectively stimulated at different depths, \ie, at $d=$ \SIlist{0.3; 0.5; 0.7; 0.9; 1.1}{\centi\meter}.
To do so, we first subdivide the muscle into two motor units. 
All recruited fibres are grouped into the first motor unit (MU1). 
The territory of MU1 is defined by all points at the cross sectional coordinates $x_\mathrm{t}^{\parallel}=\SI{0.75}{\centi\meter}$ and $x_\mathrm{t}^{\perp}=\SI{2}{\centi\meter} - d$. 
The territory of the second motor unit (MU2) contains all points that are not included in the territory of MU1. 
Hence, for both motor units, we choose $f_\mathrm{r}^k$ ($i=1,2$). 
To stimulate the fibres, a single current pulse with amplitude \SI{700}{\milli\ampere\per\square\centi\meter} and length \SI{0.1}{\milli\second} is applied to the muscle fibre membranes of MU1 at their neuromuscular junctions, \ie, at $x_\mathrm{l}^{\parallel}=\SI{1}{\centi\meter}$, $x_\mathrm{t}^{\parallel}=\SI{0.75}{\centi\meter}$ and $x_\mathrm{t}^{\perp}=\SI{2}{\centi\meter} - d$.
In order to compare measurements from the muscle surface and the body surface, the simulations are conducted for an isolated muscle (\ie, $d_\mathrm{fat}=$ \SI{0.0}{\centi\meter}) as well as with adipose tissue layers with thickness $d_\mathrm{fat}=$ \SIlist{0.2; 0.4}{\centi\meter} on top of that muscle.
All other model parameters are summarised in Table~\ref{tab:parameters}. 
Based on these parameters, the intracellular conductivity tensors are calculated by $\boldsymbol{\sigma}_\mathrm{i}^k = \sigma_\mathrm{i}^\mathrm{l} \, \boldsymbol{f} \otimes \boldsymbol{f}$ ($\forall k \in \mathscr{M}_\mathrm{MU}$), where $\boldsymbol{f}$ is a unit vector aligned with the muscle fibre direction.
Accordingly, the extracellular conductivity tensor is given by $\boldsymbol{\sigma}_\mathrm{e} = \sigma_\mathrm{e}^\mathrm{l} \, \boldsymbol{f} \otimes \boldsymbol{f} \, + \, \sigma_\mathrm{e}^\mathrm{t} \, \left(\boldsymbol{I} - \boldsymbol{f} \otimes \boldsymbol{f} \right)$ with $\boldsymbol{I}$ being the second-order identity tensor.
To simulate the behaviour of the muscle fibre membranes, we appeal to the model of \cite{Hodgkin1952}, which was imported from the models repository of the Physiome Project\footnote{https://models.physiomeproject.org/cellml} (cf. \cite{Lloyd2004}).
Finally we note that the given model can only be solved numerically and the applied methods are presented in Appendix~\ref{sec:numerics}. 

\subsection{Virtual EMG and MMG recordings and data analysis}
The computational model yields at each time step and each grid point a prediction for the electrical potential and magnetic field.
We assume an idealised recording system that does not affect the physical fields. 
It measures at a selected discrete location (\ie channel) the extracellular potential (or the body potential) and all three components of the magnetic field yielding a measurement vector 
\begin{equation}
  \boldsymbol{m}(\boldsymbol{x},t) = \begin{cases} [\phi_\mathrm{e}, B_\mathrm{l}^{\parallel}, B_\mathrm{t}^{\parallel}, B_\mathrm{t}^{\perp}]^T \ , & \boldsymbol{x} \in \mathrm{\Omega}_\mathrm{m} \ , \\ 
  [\phi_\mathrm{b}, B_\mathrm{l}^{\parallel}, B_\mathrm{t}^{\parallel}, B_\mathrm{t}^{\perp}]^T \ , & \boldsymbol{x} \in \mathrm{\Omega}_\mathrm{b} \ .
  \end{cases}
\end{equation} 
Therein $B_\mathrm{l}^{\parallel}$ is the magnetic field component aligned with the muscle fibres (and tangential to the muscle surface), $B_\mathrm{t}^{\parallel}$ is the component of the magnetic field orthogonal to the muscle fibres and tangential to the surface, and $B_\mathrm{t}^{\perp}$ is the magnetic field component normal to the body surface (and orthogonal to the muscle fibres), cf. Fig.~\ref{fig:geo}.  
We assume a sampling frequency of \SI{10000}{Hz} for both the synthetic EMG and MMG. 

To quantitatively evaluate the relation between the amplitude of the signal components and the geometrical configuration, the root-mean-square (RMS) value is calculated for the virtual EMG and MMG signals.
Further, the spectral content of the virtual signals is investigated by estimating the power spectral density (PSD). 
Both metrics provide insights on the spatial resolution of EMG and MMG signals.


\section{Results}\label{sec:results}

\subsection{Single channel recordings at the muscle surface}\label{sec:results_selectivity_intramuscular}
As baseline experiment, the spatial resolution of EMG and MMG signals is investigated for an isolated muscle. To do so, the muscle fibres are selectively stimulated in different depths within the muscle tissue (cf. Sect.~\ref{sec:in_silico_exp}). 
The muscle response is observed from a single channel, which is placed between the innervation zone and the boundary of the muscle on its surface (cf. Fig.~\ref{fig:geo}), \ie, $x_\mathrm{l}^{\parallel} = \SI{2.5}{\centi\meter}$ and $x_\mathrm{t}^{\parallel} = \SI{0.6}{\centi\meter}$. 
The bottom row of Fig.~\ref{fig:res_depth} shows that the amplitude of all components of measurement vector $\boldsymbol{m}$ (\ie, the extracellular potential $\phi_\mathrm{e}$ and three components of the magnetic field $\boldsymbol{B}$) decreases with increasing activation depth.
In detail, the decrease in amplitude is most distinct for the surface normal component of the magnetic field, \ie, for a depth of \SI{1.1}{\centi\meter} the RMS decreases by a factor of 0.019 if compared to the RMS at $d=$ \SI{0.3}{\centi\meter} (cf. Table~\ref{tab_depth}).
The signal decay is least pronounced for the magnetic field component tangential to the body surface and orthogonal to the muscle fibre direction, \ie, for a depth of \SI{1.1}{\centi\meter} the RMS decreases by a factor of 0.317 of the RMS at $d=$ \SI{0.3}{\centi\meter}.
For the same condition the RMS of the EMG decreases by a factor of 0.124 and the MMG component aligned with the muscle fibres decreases by a factor of 0.031.
Further, from Fig.~\ref{fig:res_depth} and Table~\ref{tab_depth} it can be seen that increasing the depth of the stimulated fibres causes a left-shift in the mean frequency content of the observed signals.
This indicates a spatial low-pass filtering effect of the muscle tissue, of which the surface normal component of the magnetic field exhibits the lowest modulation.
Further, the shift in the mean frequency content is relatively smaller for the EMG than for the $x_\mathrm{l}^{\parallel}$-component and the $x_\mathrm{t}^{\parallel}$-component of the MMG.

\begin{table}[h!]
\begin{tabular}{ l  c  c  c  c  c  } 
	\toprule
	Depth (\SI{}{\centi\meter})  & 0.3 & 0.5 & 0.7 & 0.9 & 1.1 \\ \toprule
	EMG-RMS  & 1 & 0.468 & 0.270 & 0.177 & 0.124 \\  
	EMG-MNF & 1 & 0.802 & 0.714 & 0.678 & 0.660 \\ \midrule
	MMG-RMS ($x_\mathrm{l}^{\parallel}$) & 1 & 0.445 & 0.184 & 0.076 & 0.031\\  
	MMG-MNF ($x_\mathrm{l}^{\parallel}$) & 1 & 0.788 & 0.699 & 0.639 & 0.592 \\ \midrule
	MMG-RMS ($x_\mathrm{t}^{\parallel}$) & 1 & 0.728 & 0.507 & 0.3766 & 0.3171\\  
	MMG-MNF ($x_\mathrm{t}^{\parallel}$) & 1 & 0.715 & 0.598 & 0.521 & 0.444\\ \midrule
	MMG-RMS ($x_\mathrm{t}^{\perp}$) & 1 & 0.286 & 0.101 & 0.041 & 0.019\\  
	MMG-MNF ($x_\mathrm{t}^{\perp}$) & 1 & 0.902 & 0.858 & 0.847 & 0.854 \\ \bottomrule
\end{tabular} 
\caption[]{Effect of the depth of the activated muscle fibres on the RMS and the MNF of the surface EMG signal and surface MMG signal. Note that for the virtual MMG recordings each component of the magnetic field is measured individually and which is indicated by the respective coordinate shown in brackets. All values are normalised with respect to the values from the simulation with the lowest depth, \ie, \SI{0.3}{\centi\meter}.}\label{tab_depth} 
\end{table}

\subsection{Single channel recordings at the body surface}\label{sec:results_selectivity_surface}
To investigate the influence of adipose tissue on non-invasively observable surface signals, we compare the computed fields for three cases with variable fat tissue thickness, \ie, \SI{0}{\centi\meter}, \SI{0.2}{\centi\meter} and \SI{0.4}{\centi\meter}.
The distance between the recording point and the active fibres is kept constant.
Hence, when a thicker fat tissue layer is simulated more superficial fibres are stimulated, \ie, $d=$ \SI{0.9}{\centi\meter}, \SI{0.7}{\centi\meter} and \SI{0.5}{\centi\meter}, respectively.
Again, the muscle's response is observed from a single channel at $x_\mathrm{l}^{\parallel} = \SI{2.5}{\centi\meter}$, and $x_\mathrm{t}^{\parallel} = \SI{0.6}{\centi\meter}$.   
Fig.~\ref{fig:res_fat_2} depicts that the amplitude of the surface signal strongly depends on the thickness of the fat tissue layer for the EMG. 
The same holds for the $x_\mathrm{t}^{\parallel}$-component and the $x_\mathrm{f}^{\parallel}$-component of the MMG.
In detail, for the \textit{in silico} experiments with fat tissue layers of \SI{0.2}{\centi\meter} and \SI{0.4}{\centi\meter}, the RMS of the EMG signal increases by a factor of 1.67 and 2.72 when compared to the case without fat.
For the MMG component aligned with the muscle fibres, the RMS decreases by a factor of 0.80 and 0.61, respectively.
As far as the $x_\mathrm{t}^{\parallel}$-component of the MMG is concerned, the RMS values change by a factor of 0.55 ($d_\mathrm{fat}=$ \SI{0.2}{\centi\meter}) and 0.63 ($d_\mathrm{fat}=$ \SI{0.4}{\centi\meter}) compared to the respective reference RMS value without fat. Thereby, one also observes a notably modulated shape of the surface potential.
This is also reflected by a change of the signal's frequency spectrum.
In contrast, the amplitude and the frequency content of the normal-to-the-body-surface component are less affected by the adipose tissue.
For the \textit{in silico} experiment with $d_\mathrm{fat}=$ \SI{0.4}{\centi\meter}, the RMS value of the $x_\mathrm{t}^{\perp}$-component changes only by a factor of 1.16 compared to the simulation without fat.

\subsection{The spatial distribution of the amplitude for surface signals}\label{sec:results_surface_distribution}
Further insights on the spatial selectivity of both EMG and MMG signals can be gained, when considering the dependency between the sensor position and the bioelectromagentic signals. 
To do so, we evaluate the root mean square (RMS) for all components of the measurement vector $\boldsymbol{m}$ in a line orthogonal to the muscle fibres and mid way through the innervation zone and the muscle boundary (cf. Fig.~\ref{fig:geo}).
Fig.~\ref{fig:spatial_rms} shows that the spatial distribution of the signal's power is fundamentally different between the EMG and the MMG-components. 
For the EMG signal, the amplitude reaches its maximal value directly over the active fibres.
For the $x_\mathrm{l}^{\parallel}$-component and $x_\mathrm{t}^{\perp}$-component of the MMG, the signal's amplitude is zero directly over the source.
Further, the depth of the active fibre correlates with the distance to the maximum.
Considering the case without fat, the distance between the zero value of the $x_\mathrm{l}^{\parallel}$-component (directly over the source) and the maximal RMS value is \SI{0.2}{\centi\meter} for a fibre depth of \SI{0.3}{\centi\meter}, \SI{0.3}{\centi\meter} for a fibre depth of \SI{0.5}{\centi\meter}, and saturates at \SI{0.35}{\centi\meter} for higher fibre depths. 
Similarly, for the $x_\mathrm{t}^{\perp}$-component and in the case without fat, the distance between the maximum RMS value and the zero value is \SI{0.2}{\centi\meter} for a fibre depth of \SI{0.3}{\centi\meter}, \SI{0.3}{\centi\meter} for a fibre depth of \SI{0.5}{\centi\meter}, \SI{0.4}{\centi\meter} for a fibre depth of \SI{0.7}{\centi\meter}, \SI{0.45}{\centi\meter} for a fibre depth of \SI{0.9}{\centi\meter} and \SI{0.5}{\centi\meter} for a fibre depth of \SI{1.1}{\centi\meter}. 
Further, it can be seen that the RMS distribution of the $x_\mathrm{t}^{\parallel}$-MMG-component strongly depends on the fat tissue layer and does not follow a distinct pattern. 
When increasing the thickness of the fat tissue layer, for the EMG it can be observed that the inter-channel variability gets strongly compressed. For example, for a fibre depth of \SI{0.3}{\centi\meter} the coefficient of variation of the RMS values is \SI{63.6}{\%} for the case without fat, \SI{43.4}{\%} for $d_\mathrm{fat}=$ \SI{0.2}{\centi\meter} and \SI{22.9}{\%} for $d_\mathrm{fat}=$ \SI{0.4}{\centi\meter}.
In contrast, the MMG components aligned with the muscle fibres and normal to the surface better preserve the inter-channel variability.
Considering the \textit{in silico} experiment with a fibre depth of \SI{0.3}{\centi\meter}, the coefficient of variation of the RMS values for the $x_\mathrm{l}^{\parallel}$-component is \SI{73.1}{\%} in the case there is no fat, \SI{61.6}{\%} for $d_\mathrm{fat}=$ \SI{0.2}{\centi\meter} and \SI{58.6}{\%} for $d_\mathrm{fat}=$ \SI{0.4}{\centi\meter}. For the $x_\mathrm{t}^{\perp}$-component the coefficient of variation of the RMS values is \SI{49.5}{\%} in the case without fat, \SI{40.2}{\%} for $d_\mathrm{fat}=$ \SI{0.2}{\centi\meter} and \SI{37.9}{\%} for $d_\mathrm{fat}=$ \SI{0.4}{\centi\meter}.

\subsection{The contribution of different domains to the magnetic field}\label{sec:results_domain_contribution}
To investigate the origin of the experimentally observable magnetic fields, the MMG recorded on the body surface is split up into the contribution of the different domains.
To do so, we exploit the linearity of the magnetic field equations, \ie, Eqn.~\ref{eq:mag_pot_muscle} and Eqn.~\ref{eq:mag_pot_body}. Therefore, the solution of the overall magnetic field problem can be linearly reconstructed from the individual solutions of each right hand term, \ie, the contribution of each domain / region (cf. Sect.~\ref{sec:modelling_magnetic_field}).
In Fig.~\ref{fig:domain_contributions} this is exemplary shown for the \textit{in silico} experiment with a fat tissue layer of \SI{0.2}{\centi\meter} and active muscle fibres in a depth of \SI{0.5}{\centi\meter}.
It can be observed that the component of the magnetic field aligned with the muscle fibres, \ie, the $x_\mathrm{l}^{\parallel}$-component, is completely determined by volume currents in the extracellular space and the body region. The RMS of the extracellular contribution is 0.950 and the RMS of the body region contribution is 0.051 (normalised with respect to the RMS value of the observable magnetic field).
In contrast, the magnetic field components orthogonal to the muscle fibre direction, \ie, the $x_\mathrm{t}^{\parallel}$-component and the $x_\mathrm{t}^{\perp}$-component, depend on currents from all domains.
Thereby, the non-recruited muscle fibres considerably contribute to the experimentally observable magnetic field; for the presented simulation, the currents in the active and passive muscle fibres have opposite directions and thus mutually limit their visibility in the observable magnetic field.
In detail, for the $x_\mathrm{t}^{\parallel}$-component the domain specific RMS values normalised with respect to the measurable field are 0.999 for the extracellular space, 0.670 for the active intracellular domains, 0.357 for the non stimulated intracellular domains and 0.027 for the body region.
Considering the normal-to-the-body-surface component, then the active fibres dominate the measurable signal, \ie, the RMS normalised with respect to the RMS value of the observable magnetic field is 1.485. Further, the normalised RMS values are 0.517 for the passive intracellular domains, 0.134 for the extracellular space and 0.002 for the body region.   

\section{Discussion}
Within this work we propose a novel \textit{in silico} framework to simulate electro-magnetic fields induced by the activity of skeletal muscles.
The model is used for the first systematic comparison between the well established EMG measurements, cf. \citet{Merletti2016}, and MMG which recently gained attention due to progress in sensor technology \citep[cf. e.g.,][]{Zuo2020,Broser2018,Broser2021,Llinas2020}. 

\subsection{Limitations}
The presented systemic multi-scale approach integrates the most important features of the microstructure, for example, the shape of the muscle fibres and the electrical behaviour of the fibre membranes. 
It must be noted that the effects of currents in complex microstructural features such as the T-tubuli system is beyond the scope of the proposed model.
Further, while we use an idealised cubic muscle geometry to ignore geometric effects and illustrate the basic properties of the magnetic field induced by active muscles, the MMG is expected to strongly depend on the specific muscle geometry.  
Thereby, it should be noted that the presented continuum field approach provides a high flexibility to resolve arbitrary muscle geometries by employing discretization schemes such as the finite element method, e.g., \citet{Heidlauf2016,Mordhorst2015,Schmid2019}. 
Finally, we note that within this work we focused on the physical properties of the bioelectromagentic fields. Hence, we considered idealised sensors that can record from a single point in space and measurements are unaffected by noise. However, for specific applications the specific sensor properties need to be considered for comparing EMG and MMG.   

\subsection{The spatial resolution of EMG and MMG}
The spatial resolution is one of the most extensively discussed property of EMG.
That is, when employing invasive intramuscular electrodes, EMG is highly sensitive with respect to the spatial coordinate of the recording point. 
However, when measured non-invasively from the skin EMG has a poor spatial resolution as surrounding electrically inactive tissues, such as, for example, fat, act as a spatial low pass filter. 
Within this work we address the hypothesis that surface MMG can overcome the limitations of surface EMG's spatial selectivity.
We did so by carrying out an \textit{in silico} comparison between both bioelectric and biomagnetic signals.

EMG and MMG measure different physical fields and therefore are not directly comparable.
Thus, as a reference experiment we investigated the spatial properties for EMG and MMG signals directly recorded on the surface of an isolated muscle.
When the distance between the recording point and the active muscle fibres is increased, all MMG components and the EMG show a strong decrease in amplitude. Hence, we conclude that for both intramuscular electrical field and magnetic field recordings the spatial resolution should be reasonable to observe local events within the tissue.
This, however, changes, if we consider non-invasive surface recordings (which are affected by electrically inactive tissues such as fat).
Our simulations show, as previously reported, \eg, \citet{Roeleveld1997,Lowery2002,Farina2002}, that the spatial selectivity of the EMG is compromised.
We conclude this from the fact that increasing the thickness of the adipose tissue causes a strong modulation of the EMG signal's amplitude (cf. Fig.~\ref{fig:res_fat_2} and Fig.~\ref{fig:spatial_rms}).

Considering the MMG's $x_\mathrm{t}^{\parallel}$-component, the effect of fat tissue on the surface signal is even more pronounced than for EMG.
However, in comparison to EMG, our simulations show that the MMG components normal to the surface and aligned with the muscle fibres exhibit a much less pronounced influence of the fat layer.
Particularly, the spatial-temporal pattern of the normal-to-the-surface component is nearly preserved (cf. Fig.~\ref{fig:res_fat_2} and Fig.~\ref{fig:spatial_rms}).     
Thus, we conclude that a careful selection of the measured magnetic field component can overcome the limitations given by the poor spatial selectivity of surface EMG.

The potentially most interesting implication of surface MMG's increased spatial selectivity is the relatively higher sparseness of the magnetic interference signals.
This advocates for the use of non-invasive MMG recordings to decode the neural drive to a muscle using, \eg \citet{Nawab2010,Holobar2010,Farina2014b,Negro2016} as the impact on the interference of the different sources will be less pronounced. 
Further, it limits the signal's contamination with cross-talk.
On the other hand, it should be noted that a higher spatial selectivity also implies that rather local properties of the muscle tissue are observed. 
This, if not compensated by a congruous amount of sensors, may compromise the robustness and comparability of measurements as a too pronounced weighting of local properties yields the risk to bias the observations.
This is a well-known limitation of intramuscular EMG decomposition  \citep[cf. \eg,][]{DeLuca2006,Farina2010,Farina2012}.
Further, it is noted that a higher spatial selectivity makes measurements more susceptible for motion artifacts. 

\subsection{The biophysical origin of the measurable magnetic field}
We make use of the systemic modelling framework to deduce the biophysical origin of the magnetic field induced by muscle activity.
An electrical current only can generate a magnetic field circular to the direction of the current.
Accordingly, we showed that the magnetic field aligned with the muscle fibres, is fully determined by volume currents in the extracellular space / surrounding tissues.
In contrast, both magnetic field components orthogonal to the muscle fibre direction contain contributions from intracellular currents, which are in the literature sometimes referred to as primary currents \citep{Malmivuo1995}.
However, while the MMG component tangential to the body surface and orthogonal to the muscle fibres, \ie, the $x_\mathrm{t}^{\parallel}$-component, is dominated by volume currents, the surface-normal component of the MMG, \ie, the $x_\mathrm{t}^{\perp}$-component, is dominated by intracellular currents.

The observation that the surface-normal component of the magnetic field strongly reflects intracellular currents and is relatively insensitive to the effect of fat, yields several potential benefits for the interpretation of experimental data.
This can be beneficial when properties on the muscle fibre level, for example, membrane fatigue, should be estimated from MMG data.
Further, when aiming to use inverse modelling and MMG to reconstruct the sources of the bioelectromagnetic activity, \eg, \cite{Llinas2020}, a field component which is (nearly) invariant with respect to volume currents can reduce the uncertainty associated with the required estimate for the tissue's conductive properties. 
We conclude this discussion by noting that the model predicts a surprisingly big contribution of passive muscle fibres.

\subsection{Conclusion and Outlook}
Within this work we propose a systemic multi-scale model to simulate EMG and MMG. 
We show that non-invasive MMG can overcome the limitations of surface EMG, in particular with respect to its poor spatial selectivity. 
In the future, we want to use the presented modelling framework to investigate the potential of non-invasive MMG to decode the neural drive to muscles.
Further, given the emerging progress in MMG sensor technology, the presented systemic simulation framework provides excellent capabilities to assist the interpretation of experimental data as well as assisting the optimisation of MMG sensor arrays.

\appendix
\section{Numerical treatment}\label{sec:numerics}
The mathematical modelling framework presented in this work, \ie, Sect.~\ref{sec:modelling_framework}, can only be solved numerically.
The applied methods are outlined in the following.
In summary, we exploit the quasi-static conditions and decouple the electric and the magnetic field (cf. Sect.~\ref{sec:maxwells_eqs}). 
Thus, we appeal to a staggered solution scheme where (i) the electric field equations are solved for the muscle and body domain (cf. \ref{sec:numerics_multi-domain}) and (ii) the magnetic field predictions are based on the previously calculated electric potentials (cf. \ref{sec:numerics_magnetic}).
Therefore, the continuous domains are represented by a finite number of grid points and the spatial derivatives are approximated by finite differences.  
The whole model is implemented in MATLAB (The MathWorks, Inc., Natick, Massachusetts, United States) and the corresponding code is hosted on a freely accessible git repository\footnote{https://bitbucket.org/klotz\_t/multi\_domain\_fd\_code}.

\subsection{Solving for the electrical potential fields}\label{sec:numerics_multi-domain}
Given the reaction-diffusion characteristic of Eqn.~\eqref{intracellular_md_eqs}, a first-order Godunov-type splitting scheme is applied to yield
\begin{subequations}
\begin{align}
  \frac{V_\mathrm{m}^{k,t_*}-V_\mathrm{m}^{k,t_i}}{\Delta t} \ &= \ - \frac{1}{C_\mathrm{m}^k} \, I_\mathrm{ion}^k(\boldsymbol{y}^k,V_\mathrm{m}^k) \ ,   \label{eq:split_reac_term} \\
  \frac{V_\mathrm{m}^{k,t_{i+1}}-V_\mathrm{m}^{k,t_*}}{\Delta t} \ &= \ \frac{1}{C_\mathrm{m}^k A_\mathrm{m}^k} \, \Big(
  \operatorname{div} \left[\boldsymbol{\sigma}_\mathrm{i}^k \operatorname{grad} V_\mathrm{m}^k \right] \, + & \notag \\  
  & \qquad \qquad \ + \, \operatorname{div} \left[\boldsymbol{\sigma}_\mathrm{i}^k \operatorname{grad} \phi_\mathrm{e} \right] \Big) \ ,  \label{eq:split_diffusion}
\end{align}
\end{subequations}
for each skeletal muscle material point $\mathcal{P}\in \mathrm{\Omega_m}$ and $k \in \mathscr{M}_\mathrm{MU}$.
Therein, a first-order forward finite difference is used to approximate the temporal derivatives and $t_*$ denotes an intermediate time step.
Note that the hereby introduced splitting error becomes acceptable for sufficiently small time steps.
Herein, we chose as time step $\Delta t = \SI{0.1}{\milli\second}$.
Further, note that both Eqn.~\eqref{eq:split_reac_term} and Eqn.~\eqref{eq:split_diffusion} are still continuous in space.
This allows us to use specialised solution schemes for the reactive and the diffusive parts of the model.

In detail, Eqn.~\eqref{eq:split_reac_term} together with Eqn.~\eqref{intracellular_md_eqs_odes} forms a system of stiff ordinary differential equations, which is solved for the interval $[t_i, t_*]$ by an improved Euler method and a fixed time step of $\Delta t_\mathrm{ode} = \SI{0.001}{\milli\second}$  (cf. \cite{Bradley2018}).
Further, the coupled diffusion problem given by given by Eqn.~\eqref{extracellular_md_eqs}, Eqn.~\ref{eq:split_diffusion} and Eqn.~\eqref{eq:body_pot}, is addressed by evaluating Eqn.~\eqref{extracellular_md_eqs}, Eqn.~\eqref{eq:body_pot} and the right-hand side of Eqn.~\eqref{eq:split_diffusion} at $t=t_{i+1}$, \ie, employing an implicit Euler method, whereby the spatial derivatives are approximated with second-order accurate central finite differences. 
Accordingly, the flux boundary conditions, \ie, Eqn.~\eqref{eq:bc_Vm}, Eqn.~\eqref{eq:bc_phi_b}, Eqn.~\eqref{eq:bc_phi_e_out} and Eqn.~\eqref{eq:bc_phi_flux}, are also evaluated at $t=t_{i+1}$, while being approximated with second-order accurate forward/backward finite differences.
For the spatial discretisation, we chose equally spaced grid points and a step size of $h=\SI{0.05}{\centi\meter}$.
This discretisation yields a linear system of equations which is solved with matlab's built-in GMRES function \citep{Saad1986}.
The linear system is preconditioned via an incomplete LU factorisation (crout version, drop tolerance: 1e-6) and the following  solver options are applied: an absolute and relative tolerance of 1e-10, restart after 20 inner iterations and a maximum number of 20 outer iterations. 

\subsection{Solving for the magnetic vector potential}\label{sec:numerics_magnetic}
Based on the solution of the multi-domain model, \ie, the electrical potential fields for each time step and each grid point of the muscle region and body region, second-order central finite differences are used to obtain estimates for the (first) spatial derivative of the electrical potentials of the right-hand side of Eqn.~\eqref{eq:mag_pot_muscle} and Eqn.~\eqref{eq:mag_pot_body}.
Further, the (second) spatial derivatives given on the left-hand side of Eqn.~\eqref{eq:mag_pot_muscle} and Eqn.~\eqref{eq:mag_pot_body} are discretised using second-order central finite differences.

As for the electrical potential fields are concerned, the flux boundary conditions of the magnetic vector potential in the muscle and body region (cf. Eqn.~\eqref{eq:interface_flux_mag_pot}) are approximated with second-order accurate forward/backward finite differences.
The surrounding air is represented by an infinitely long (virtual) boundary element.
Its normal derivative (cf. Eqn.~\eqref{eq:interface_flux_mag_pot}) is approximated by a first-order forward/backward finte difference.
Recalling that the magnetic vector potential is zero far away from the bioelectromagentic sources (cf. Eqn.~\eqref{eq:zero_dirichlet_mag_pot}), the normal derivative of the magnetic vector potential vanishes at the interface between the body and air.  

In summary, this discretisation yields a linear system of equations, which needs to be solved to obtain the magnetic vector potential at each grid point of the muscle  and body region. 
This linear system is solved with MATLAB's build in ''mldivide'' function, as this function can handle multiple pre-computed right-hand side vectors simultaneously.
Finally, the magnetic field $\boldsymbol{B}$ is calculated via Eqn.~\eqref{eq:mag_vec_pot_def} in a post-processing step. 
Thereby the (first) spatial derivatives of the curl operator are approximated by second-order central finite differences.

\begin{acknowledgements}
This research was funded by the Deutsche Forschungsgemeinschaft (DFG, German Research Foundation) under Germany’s Excellence Strategy EXC 2075 390740016.
\end{acknowledgements}

\bibliographystyle{spbasic}      
\bibliography{bibliography}   

\begin{thebibliography}{43}
\providecommand{\natexlab}[1]{#1}
\providecommand{\url}[1]{{#1}}
\providecommand{\urlprefix}{URL }
\expandafter\ifx\csname urlstyle\endcsname\relax
  \providecommand{\doi}[1]{DOI~\discretionary{}{}{}#1}\else
  \providecommand{\doi}{DOI~\discretionary{}{}{}\begingroup
  \urlstyle{rm}\Url}\fi
\providecommand{\eprint}[2][]{\url{#2}}

\bibitem[{Bradley et~al(2018)Bradley, Emamy, Ertl, G\"oddeke, Hessenthaler,
  Klotz, Kr\"amer, Krone, Maier, Mehl, Rau, and R\"ohrle}]{Bradley2018}
Bradley CP, Emamy N, Ertl T, G\"oddeke D, Hessenthaler A, Klotz T, Kr\"amer A,
  Krone M, Maier B, Mehl M, Rau T, R\"ohrle O (2018) Enabling detailed,
  biophysics-based skeletal muscle models on hpc systems. Frontiers in
  Physiology 9:816, \doi{10.3389/fphys.2018.00816}

\bibitem[{Broser et~al(2018)Broser, Knappe, Kajal, Noury, Alem, Shah, and
  Braun}]{Broser2018}
Broser PJ, Knappe S, Kajal DS, Noury N, Alem O, Shah V, Braun C (2018)
  Optically pumped magnetometers for magneto-myography to study the innervation
  of the hand. IEEE Transactions on Neural Systems and Rehabilitation
  Engineering 26(11):2226--2230

\bibitem[{Broser et~al(2021)Broser, Middelmann, Sometti, and
  Braun}]{Broser2021}
Broser PJ, Middelmann T, Sometti D, Braun C (2021) Optically pumped
  magnetometers disclose magnetic field components of the muscular action
  potential. Journal of Electromyography and Kinesiology 56:102,490

\bibitem[{Bryant(1969)}]{Bryant1969}
Bryant SH (1969) Cable properties of external intercostal muscle fibres from
  myotonic and nonmyotonic goats. The Journal of Physiology 204:539 -- 550,
  \doi{10.1113/jphysiol.1969.sp008930}

\bibitem[{Cohen and Givler(1972)}]{Cohen1972}
Cohen D, Givler E (1972) Magnetomyography: Magnetic fields around the human
  body produced by skeletal muscles. Applied Physics Letters 21(3):114--116

\bibitem[{De~Luca et~al(2006)De~Luca, Adam, Wotiz, Gilmore, and
  Nawab}]{DeLuca2006}
De~Luca CJ, Adam A, Wotiz R, Gilmore LD, Nawab SH (2006) Decomposition of
  surface emg signals. Journal of neurophysiology 96(3):1646--1657

\bibitem[{Dimitrova et~al(1999)Dimitrova, Dimitrov, and
  Dimitrov}]{Dimitrova1999}
Dimitrova NA, Dimitrov AG, Dimitrov GV (1999) Calculation of extracellular
  potentials produced by an inclined muscle fibre at a rectangular plate
  electrode. Medical Engineering \& Physics 21:583--588,
  \doi{10.1016/S1350-4533(99)00087-9}

\bibitem[{Farina and Negro(2012)}]{Farina2012}
Farina D, Negro F (2012) Accessing the neural drive to muscle and translation
  to neurorehabilitation technologies. IEEE Reviews in biomedical engineering
  5:3--14

\bibitem[{Farina et~al(2002)Farina, Cescon, and Merletti}]{Farina2002}
Farina D, Cescon C, Merletti R (2002) Influence of anatomical, physical, and
  detection-system parameters on surface emg. Biological cybernetics
  86(6):445--456

\bibitem[{Farina et~al(2004)Farina, Mesin, and Martina}]{Farina2004b}
Farina D, Mesin L, Martina S (2004) Advances in surface electromyographic
  signal simulation with analytical and numerical descriptions of the volume
  conductor. Medical \& Biological Engineering \& Computing 42:467--476,
  \doi{10.1007/BF02350987}

\bibitem[{Farina et~al(2010)Farina, Holobar, Merletti, and Enoka}]{Farina2010}
Farina D, Holobar A, Merletti R, Enoka RM (2010) Decoding the neural drive to
  muscles from the surface electromyogram. Clinical neurophysiology
  121(10):1616--1623

\bibitem[{Farina et~al(2014)Farina, Merletti, and Enoka}]{Farina2014b}
Farina D, Merletti R, Enoka RM (2014) The extraction of neural strategies from
  the surface emg: an update. Journal of Applied Physiology 117(11):1215--1230

\bibitem[{Griffiths(2013)}]{Griffiths2013}
Griffiths DJ (2013) {Introduction to electrodynamics; 4th ed.} Pearson, Boston,
  MA

\bibitem[{Heckman and Enoka(2012)}]{Heckman2012}
Heckman CJ, Enoka RM (2012) Motor {U}nit. Comprehensive Physiology 2:2629--2682

\bibitem[{Heidlauf et~al(2016)Heidlauf, Klotz, Altan, Bleiler, Siebert, Rode,
  and Röhrle}]{Heidlauf2016}
Heidlauf T, Klotz T, Altan E, Bleiler C, Siebert T, Rode C, Röhrle O (2016) A
  multi-scale continuum model of skeletal muscle mechanics predicting force
  enhancement based on actin-titin interaction. Biomechanics and Modeling in
  Mechanobiology 11(10):1424 -- 1437, \doi{10.1007/s10237-016-0772-7}

\bibitem[{Hodgkin and Huxley(1952)}]{Hodgkin1952}
Hodgkin AL, Huxley AF (1952) A quantitative description of membrane current and
  its application to conduction and excitation in nerve. The Journal of
  Physiology 117(4):500--544, \doi{10.1113/jphysiol.1952.sp004764}

\bibitem[{Holobar et~al(2010)Holobar, Minetto, Botter, Negro, and
  Farina}]{Holobar2010}
Holobar A, Minetto MA, Botter A, Negro F, Farina D (2010) Experimental analysis
  of accuracy in the identification of motor unit spike trains from
  high-density surface emg. IEEE Transactions on Neural Systems and
  Rehabilitation Engineering 18(3):221--229

\bibitem[{Keener and Sneyd(2009)}]{Keener2009b}
Keener J, Sneyd J (2009) Mathematical {P}hysiology {I}{I}: {C}ellular
  {P}hysiology, vol~2, 2nd edn. Springer

\bibitem[{Klotz et~al(2020)Klotz, Gizzi, Yavuz, and R{\"o}hrle}]{Klotz2020}
Klotz T, Gizzi L, Yavuz U, R{\"o}hrle O (2020) Modelling the electrical
  activity of skeletal muscle tissue using a multi-domain approach.
  Biomechanics and modeling in mechanobiology 19(1):335--349

\bibitem[{Llin{\'a}s et~al(2020)Llin{\'a}s, Ustinin, Rykunov, Walton, Rabello,
  Garcia, Boyko, and Sychev}]{Llinas2020}
Llin{\'a}s RR, Ustinin M, Rykunov S, Walton KD, Rabello GM, Garcia J, Boyko A,
  Sychev V (2020) Noninvasive muscle activity imaging using magnetography.
  Proceedings of the National Academy of Sciences 117(9):4942--4947

\bibitem[{Lloyd et~al(2004)Lloyd, Halstead, and Nielsen}]{Lloyd2004}
Lloyd CM, Halstead MD, Nielsen PF (2004) Cellml: its future, present and past.
  Progress in Biophysics and Molecular Biology 85(2):433 -- 450,
  \doi{https://doi.org/10.1016/j.pbiomolbio.2004.01.004}, modelling Cellular
  and Tissue Function

\bibitem[{Lowery et~al(2002)Lowery, Stoykov, Taflove, and Kuiken}]{Lowery2002}
Lowery MM, Stoykov NS, Taflove A, Kuiken TA (2002) {A Multiple-Layer
  Finite-Element Model of the Surface EMG Signal}. IEEE Transactions on
  Biomedical Engineering 49(5):446--454, \doi{10.1109/10.995683}

\bibitem[{MacIntosh et~al(2006)MacIntosh, Gardiner, and
  McComas}]{MacIntosh2006}
MacIntosh R B, Gardiner F P, McComas J A (2006) Skeletal Muscle: Form and
  Function, 2nd edn. Human Kinetics

\bibitem[{Malmivuo et~al(1995)Malmivuo, Plonsey et~al}]{Malmivuo1995}
Malmivuo J, Plonsey R, et~al (1995) Bioelectromagnetism: principles and
  applications of bioelectric and biomagnetic fields. Oxford University Press,
  USA

\bibitem[{Merletti and Farina(2016)}]{Merletti2016}
Merletti R, Farina D (2016) Surface electromyography: physiology, engineering,
  and applications. John Wiley \& Sons

\bibitem[{Mesin(2005)}]{Mesin2005}
Mesin L (2005) {Analytical Generation Model Of Surface Electromyogram For
  Multi-layer Volume Conductors}. Modelling in Medicine and Biology VI, WIT
  8:95--110, \doi{10.2495/BIO050101}

\bibitem[{Mesin(2013)}]{Mesin2013}
Mesin L (2013) Volume conductor models in surface electromyography:
  Computational techniques. Computers in Biology and Medicine 43(7):942 -- 952,
  \doi{https://doi.org/10.1016/j.compbiomed.2013.02.002}

\bibitem[{Mesin et~al(2006)Mesin, Joubert, Hanekom, Merletti, and
  Farina}]{Mesin2006a}
Mesin L, Joubert M, Hanekom T, Merletti R, Farina D (2006) {A Finite Element
  Model for Describing the Effect of Muscle Shortening on Surface EMG}. IEEE
  Transactions on Biomedical Engineering 53:693--600,
  \doi{10.1109/TBME.2006.870256}

\bibitem[{Mordhorst et~al(2015)Mordhorst, Heidlauf, and
  R\"ohrle}]{Mordhorst2015}
Mordhorst M, Heidlauf T, R\"ohrle O (2015) Predicting electromyographic signals
  under realistic conditions using a multiscale chemo-electro-mechanical finite
  element model. Interface Focus 5(2):1--11, \doi{10.1098/rsfs.2014.0076}

\bibitem[{Mordhorst et~al(2017)Mordhorst, Strecker, Wirtz, Heidlauf, and
  R\"ohrle}]{Mordhorst2017}
Mordhorst M, Strecker T, Wirtz D, Heidlauf T, R\"ohrle O (2017) {POD-DEIM}
  reduction of computational {EMG} models. Journal of Computational Science
  19:86--96, \doi{10.1016/j.jocs.2017.01.009}

\bibitem[{Nawab et~al(2010)Nawab, Chang, and De~Luca}]{Nawab2010}
Nawab SH, Chang SS, De~Luca CJ (2010) High-yield decomposition of surface emg
  signals. Clinical neurophysiology 121(10):1602--1615

\bibitem[{Negro et~al(2016)Negro, Muceli, Castronovo, Holobar, and
  Farina}]{Negro2016}
Negro F, Muceli S, Castronovo AM, Holobar A, Farina D (2016) Multi-channel
  intramuscular and surface emg decomposition by convolutive blind source
  separation. Journal of neural engineering 13(2):026,027

\bibitem[{Oschman(2002)}]{Oschman2002}
Oschman JL (2002) Clinical aspects of biological fields: an introduction for
  health care professionals. Journal of Bodywork and Movement Therapies
  6(2):117--125

\bibitem[{Pullan et~al(2005)Pullan, Buist, and Cheng}]{Pullan2005}
Pullan AJ, Buist ML, Cheng LK (2005) Mathematically Modelling the Electrical
  Activity of the Heart: From Cell to Body Surface and Back Again. World
  Scientific Publishing Company, Singapore, \doi{10.1142/5859}

\bibitem[{Reincke(1993)}]{Reincke1993}
Reincke M (1993) Magnetomyographie mit dem squid - magnetomyography with the
  squid. Biomedical Engineering -- Biomedizinische Technik 38(11):276--281,
  \doi{doi:10.1515/bmte.1993.38.11.276},
  \urlprefix\url{https://doi.org/10.1515/bmte.1993.38.11.276}

\bibitem[{Roeleveld et~al(1997)Roeleveld, Blok, Stegeman, and
  Van~Oosterom}]{Roeleveld1997}
Roeleveld K, Blok J, Stegeman D, Van~Oosterom A (1997) Volume conduction models
  for surface emg; confrontation with measurements. Journal of Electromyography
  and Kinesiology 7(4):221--232

\bibitem[{R\"ohrle et~al(2019)R\"ohrle, Yavuz, Klotz, Negro, and
  Heidlauf}]{Roehrle2019}
R\"ohrle O, Yavuz U, Klotz T, Negro F, Heidlauf T (2019) Multiscale modelling
  of the neuromuscular system: coupling neurophysiology and skeletal muscle
  mechanics. Wiley Interdisciplinary Reviews: Systems Biology and Medicine

\bibitem[{Rush et~al(1963)Rush, Abildskov, and McFee}]{Rush1963}
Rush S, Abildskov J, McFee R (1963) Resistivity of body tissues at low
  frequencies. Circulation research 12(1):40--50

\bibitem[{Saad and Schultz(1986)}]{Saad1986}
Saad Y, Schultz MH (1986) Gmres: A generalized minimal residual algorithm for
  solving nonsymmetric linear systems. SIAM Journal on Scientific and
  Statistical Computing 7(3):856--869, \doi{10.1137/0907058}

\bibitem[{Schmid et~al(2019)Schmid, Klotz, Siebert, and Röhrle}]{Schmid2019}
Schmid L, Klotz T, Siebert T, Röhrle O (2019) Characterization of
  electromechanical delay based on a biophysical multi-scale skeletal muscle
  model. Frontiers in Physiology 10(1270):1 -- 13,
  \doi{10.3389/fphys.2019.01270}

\bibitem[{Woosley et~al(1985)Woosley, Roth, and Wikswo~Jr}]{Woosley1985}
Woosley JK, Roth BJ, Wikswo~Jr JP (1985) The magnetic field of a single axon: A
  volume conductor model. Mathematical Biosciences 76(1):1--36

\bibitem[{Zuo et~al(2020)Zuo, Heidari, Farina, and Nazarpour}]{Zuo2020}
Zuo S, Heidari H, Farina D, Nazarpour K (2020) Miniaturized magnetic sensors
  for implantable magnetomyography. Advanced Materials Technologies
  5(6):2000,185

\bibitem[{Zuo et~al(2021)Zuo, Nazarpour, Farina, Broser, and Heidari}]{Zuo2021}
Zuo S, Nazarpour K, Farina D, Broser P, Heidari H (2021) Modelling and analysis
  of magnetic fields from skeletal muscle for valuable physiological
  measurements. arXiv preprint arXiv:210402036

\end{thebibliography}


\begin{figure*}[h!] 
  \center
  \begin{minipage}{1.0\textwidth}
    \input{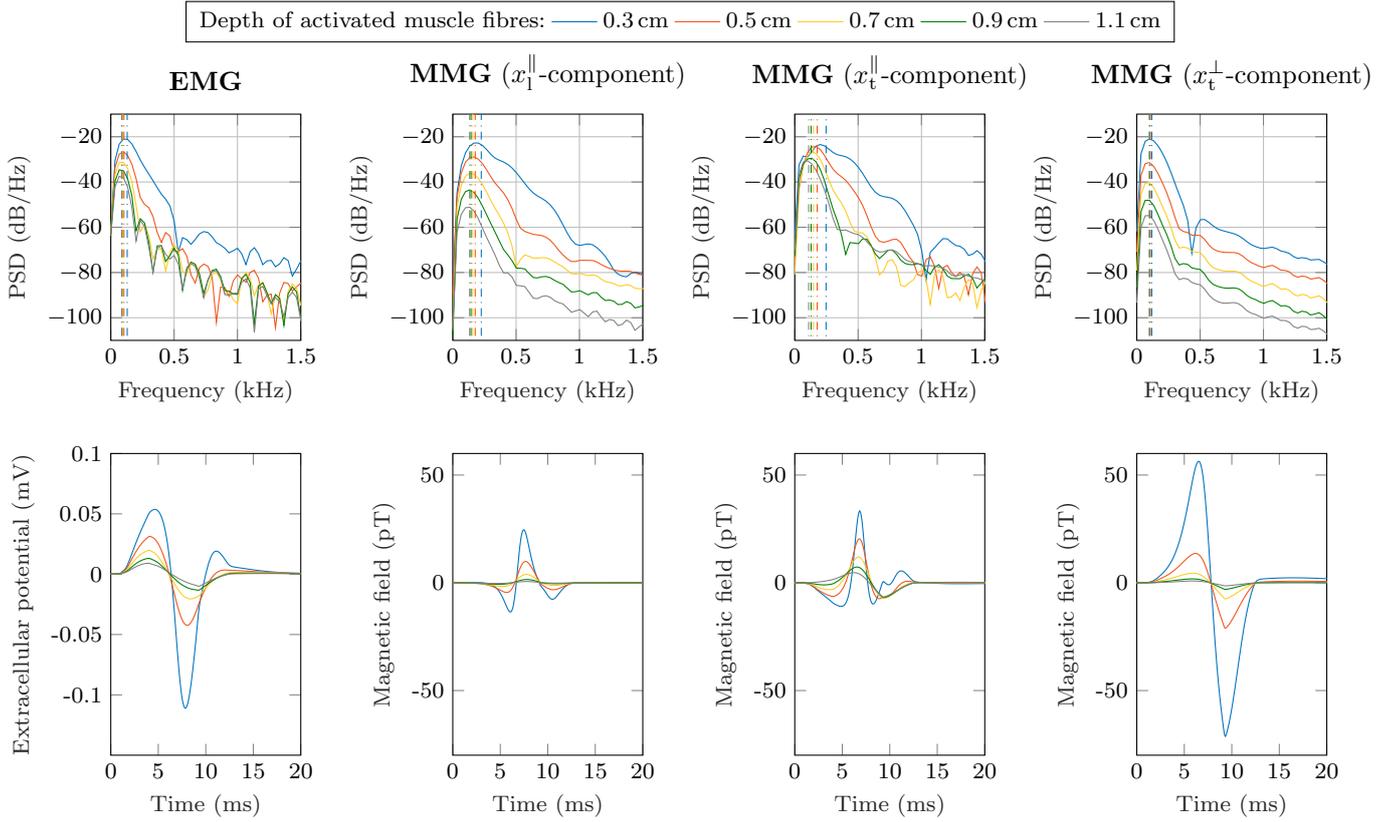}
  \end{minipage}
  \caption[]{Power spectral density (PSD) and time domain graph of the simulated surface EMG signal and surface MMG signal for variable depths of the activated muscle tissue (blue: \SI{0.3}{\centi\meter}; red: \SI{0.5}{\centi\meter}; yellow: \SI{0.7}{\centi\meter}; green: \SI{0.9}{\centi\meter}; grey: \SI{1.1}{\centi\meter}) at an arbitrary chosen virtual recording point ($x_1 = \SI{3}{\centi\meter}$, $x_2 = \SI{0.6}{\centi\meter}$). 
  Each spectrum is normalised with the total power of the respcetive EMG/MMG signal obtained from the simulation with a depth of \SI{0.3}{\centi\meter}. 
  The dashed lines in the power spectrum (top row) indicate the mean frequency content of the signal.}\label{fig:res_depth}
\end{figure*}
\begin{figure*}[h!] 
  \center
  \begin{minipage}{1.0\textwidth}
    \input{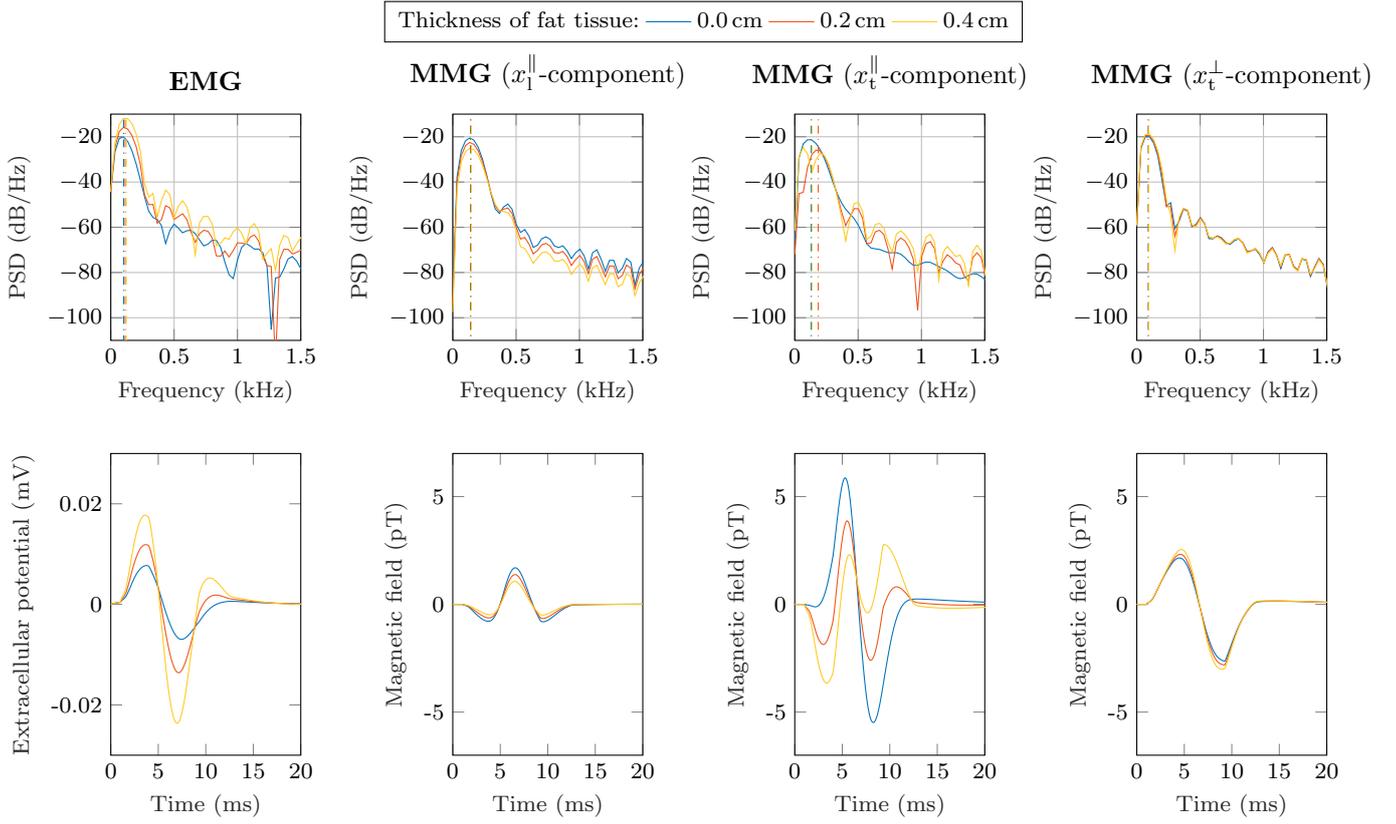}
  \end{minipage}
  \caption[]{Power spectral density (PSD) and time domain graph of the simulated surface EMG and surface MMG signal for variable thicknesses of a superficial adipose tissue layer while keeping a constant distance between active fibres and measurement location $x_\mathrm{l}^{\parallel} = \SI{2.5}{\centi\meter}$, $x_\mathrm{t}^{\parallel} = \SI{0.6}{\centi\meter}$ (in blue: \SI{0.0}{\centi\meter}; red: \SI{0.2}{\centi\meter}; yellow: \SI{0.4}{\centi\meter}).
  Each spectrum is normalised by the total power of the respective EMG/MMG signal obtained from the simulation with no fat. 
  The dashed lines in the power spectrum (top row) indicate the mean frequency content of the signal.}\label{fig:res_fat_2}
\end{figure*}
\begin{figure*}[h!]
  \center
  \input{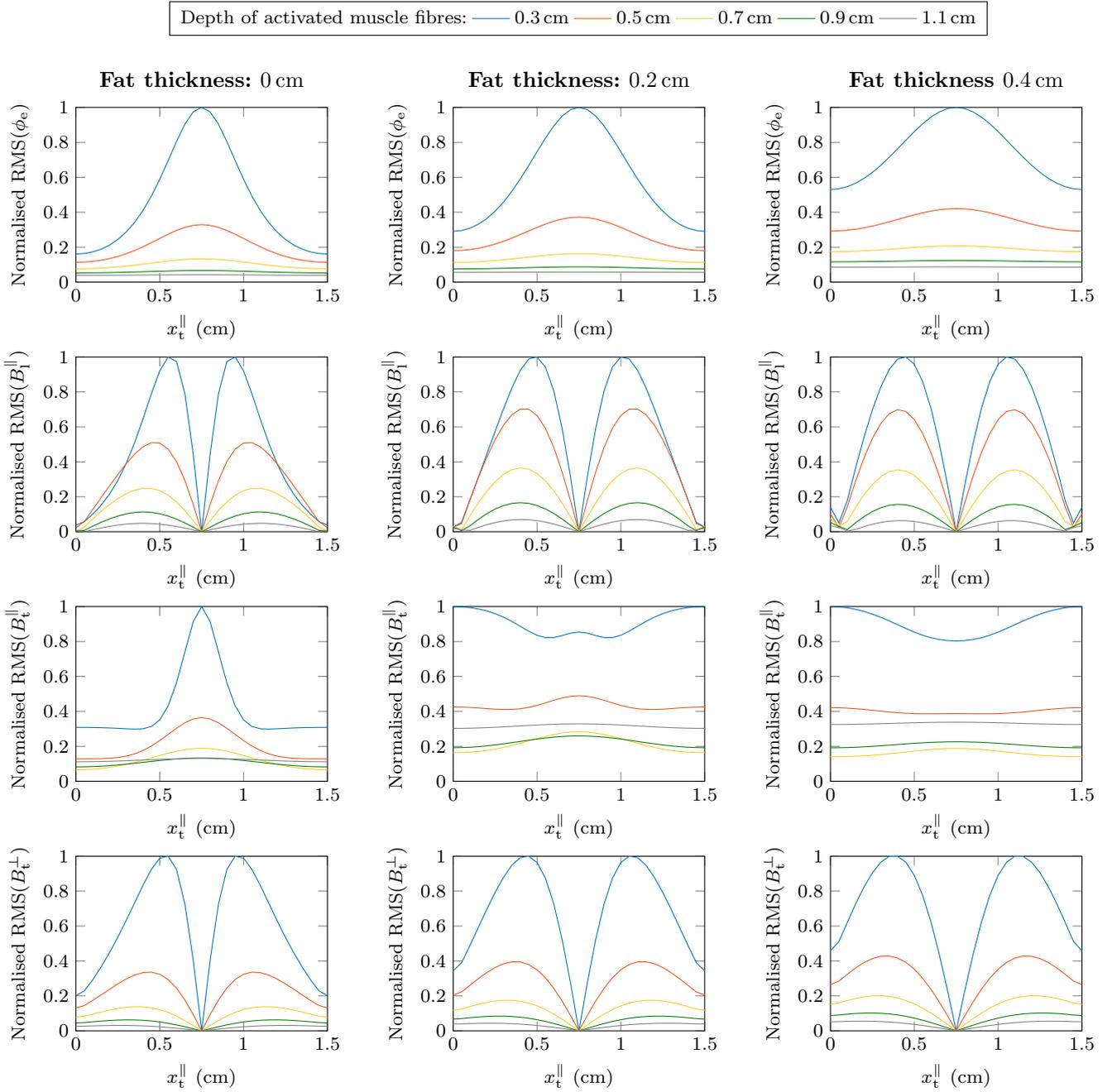}
  \caption[]{Normalised RMS values for differently thick fat tissue layers. The measurements are in a line orthogonal to the muscle fibres between the innervation zone and the muscle boundary, \ie, $x_\mathrm{l}^{\parallel} = \SI{2.5}{\centi\meter}$ on the surface of the tissue sample. From left to right the thickness of the fat tissue increases. The first row shows the EMG, the second row shows the $x_\mathrm{l}^{\parallel}$-component of the MMG, the third row shows the $x_\mathrm{t}^{\parallel}$-component of the MMG and the forth row shows $x_\mathrm{t}^{\perp}$-component of the MMG. Further, the colours indicate the depth of the activated muscle fibres (blue: \SI{0.3}{\centi\meter}; red: \SI{0.5}{\centi\meter}; yellow: \SI{0.7}{\centi\meter}; green: \SI{0.9}{\centi\meter}; grey: \SI{1.1}{\centi\meter}).}\label{fig:spatial_rms}
\end{figure*}

%
%

%
\begin{figure*}[h!]
  \center
  \input{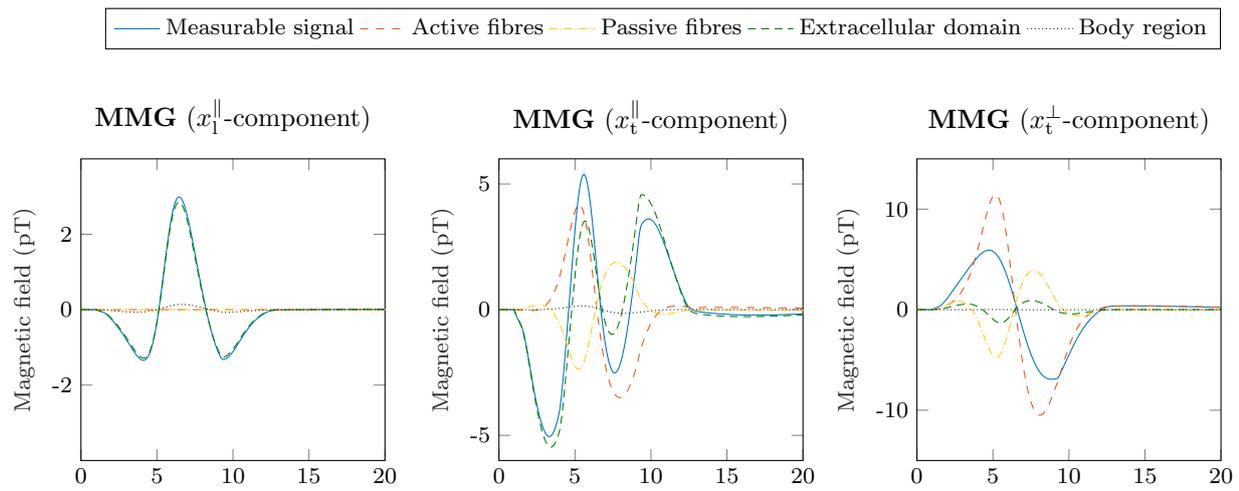}
  \caption[]{Contribution of different domains to the (magnetic) muscle action potential when selectively stimulating muscle fibres at $x_\mathrm{l}^{\parallel} = \SI{1}{\centi\meter}$, $x_\mathrm{t}^{\parallel}= \SI{0.75}{\centi\meter}$ and $x_\mathrm{l}^{\perp} = \SI{1.5}{\centi\meter}$. The virtual sensor is placed at an arbitrary point on the muscle surface ($x_\mathrm{l}^\mathrm{\parallel} = \SI{2.5}{\centi\meter}$, $x_\mathrm{t}^{\parallel} = \SI{0.6}{\centi\meter}$).}\label{fig:domain_contributions}
\end{figure*}

%

\end{document}